\documentclass[12pt, a4paper]{article}
\usepackage{tikz}
\usepackage[english]{babel}
\usepackage{amsmath}
\usepackage{amssymb}
\usepackage{enumerate}
\usepackage{amsthm}
\usepackage{eufrak}
\usepackage{graphicx}
\usepackage{verbatim}
\usepackage[hyperfootnotes=false]{hyperref}

\usepackage{cancel}

\newtheorem{thm}{Theorem}[section]
\newtheorem{prop}[thm]{Proposition}
\newtheorem{lem}[thm]{Lemma}
\newtheorem{cor}[thm]{Corollary}
\newtheorem{definition}[thm]{Definition}

\newcommand{\R}{\mathbb{R}}

\newcommand{\N}{\mathbb{N}}

\newcommand{\de}{\delta}
\newcommand{\al}{\alpha}
\newcommand{\be}{\beta}
\newcommand{\ga}{\gamma}
\newcommand{\ep}{\epsilon}
\newcommand{\ka}{\kappa}
\newcommand{\si}{\sigma}
\newcommand{\la}{\lambda}
\newcommand{\e}{\textrm{e}}

\newcommand{\Lie}{\mathcal{L}}
\newcommand{\order}{\mathcal{O}}

\DeclareMathOperator{\tr}{tr}
\DeclareMathOperator{\divergence}{div}
\DeclareMathOperator{\grad}{grad}
\DeclareMathOperator{\curl}{curl}

\begin{document}

\begin{center}

\LARGE Stationarity of asymptotically flat non-radiating electrovacuum spacetimes.
\end{center}

\begin{center}
 \setcounter{footnote}{1}

Rosemberg Toala Enriquez\footnote{\textit{Supported by CONACYT, Mexico. Grant No. 323368.}} \\
Department of Mathematics \\
University of Warwick \\
Coventry, UK.
\end{center}


\tableofcontents
\newpage

\abstract{It is proven that a solution to the Einstein-Maxwell equations whose gravitational and electromagnetic radiation fields vanish is in fact stationary in a neighbourhood of spatial infinity. That is, if the Weyl and Faraday tensors decay suitably fast, then there exists a time-like Killing vector field in the region outside the bifurcate horizon of a sphere of sufficiently large radius. In particular, truly dynamical time-periodic electrovacuum spacetimes do not exist. This is an extension of earlier work by \cite{AS} and \cite{BST} to include electromagnetism.} 

\section{Introduction}

The work presented here, roughly speaking, aims at proving a version of the folklore statement: Gravitational waves carry energy away from an isolated system. More precisely, a system with no radiation emitted to infinity must in fact be stationary. On physical grounds it is expected that truly time-periodic gravitational systems do not exist. This is due to the hyperbolic structure of the Einstein equations; any dynamical solution loses energy through outward radiation. However as remarked by Alexakis-Schlue \cite{AS}, this is a subtle mathematical question. It is true at the linearised level when we consider, for example, the free wave equation $ \square \phi=0$ on a Minkowski background: Formally, an outgoing wave with vanishing radiation field has to be stationary; that is, if the function $\phi$ decays faster than $r^{-1}$ in the null-outgoing direction then it must be time-independent \cite{Friedlander}. This is no longer the case for a suitably perturbed wave operator $L = \square + V$ as remarked by Alexakis, Schlue and Shao in \cite{Alexakis-Shao}. This problem can also be phrased as a question of uniqueness of solutions for differential operators given boundary conditions. H{\"o}rmander, \cite{Hormander}, provided general conditions for the uniqueness property to hold across an hypersurface. In the context of an hyperbolic principal symbol, these conditions reduce to requiring the hypersurface to be pseudo-convex (see section \ref{section:Carleman estimates} for more details). Complementary to this is the work of Alinhac, \cite{Alinhac}, he showed that (generically) if one of H{\"o}rmander's conditions is violated then non-uniqueness of solutions across an hypersurface ensue. \\

In this paper we focus on the consequences of the lack of gravitational waves and radiation. Firstly, we give a brief account of their mathematical meaning. It is easier to start at the linear level: Consider a small perturbation of the Minkowski metric in flat spacetime. The perturbation then is required to satisfy the linearised Einstein equations around the trivial solution. After imposing suitable gauge conditions, it is found that the perturbation obeys a wave equation and therefore it behaves as a wave propagating causally on the background. Moreover, if we impose vanishing boundary conditions at infinity then the leading order term can be interpreted as power radiated to infinity. These perturbations are the so-called \emph{gravitational waves}. The space of solutions, of both linear Einstein and gauge equations, consists of the linear combination of two scalar waves, that is, gravitational waves have 2 degrees of freedom or polarisation modes. 	\\


Now, a description of gravitational waves in the non-linear case is more delicate. To the author's knowledge there is no way of splitting the gravity field into ``stationary'' and ``dynamic'' parts; this is due to the non-linear nature of Einstein equations. Whence the lack of meaning of the expression: Here is a gravitational wave (to be attributed to the dynamical part) propagating in a background (the stationary part). Nevertheless considerable efforts have been made to understand this statement and it is possible to make sense of it \emph{at infinity} by imposing \textit{asymptotically flat} boundary conditions\footnote{Alternatively, one could impose de Sitter or anti-de Sitter boundary conditions.}. In Section \ref{section:CK analysis} we review briefly the conclusions of the asymptotic analysis carried out by Christodoulou and Klainerman in order to prove the non-linear stability of Minkowski spacetime, \cite{CK}.\\

To the author's knowledge, it was Papapetrou, \cite{Papapetrou62},\cite{Papapetrou65}, who initiated the study of the relation radiation-stationarity for the full non-linear Einstein equations. Although his result is not conclusive, it provided strong evidence for the validity of the theorem presented here. More precisely, Papapetrou showed the incompatibility of the following two conditions: a) A spacetime is stationary below a characteristic hypersurface and non-radiative above it; b) There is a shock wave along that characteristic hypersurface or above it. \\

Later on, Bi\v{c}\'ak, Sholtz and Tod, \cite{BST}, gave a rigorous proof in the class of analytic metrics using ideas from Gibbons and Stewart, \cite{Gibbons-Stewart}. Bi\v{c}\'ak-Scholtz-Tod used the (undesired) hypothesis of analyticity all the way up to infinity to split Einstein equations order by order at infinity and then used the non-radiative condition to conclude time-independence to all orders inductively. Then the desired stationarity conclusion in the interior follows by analytic continuation. \\

The theorem was later improved to cover smooth metrics by Alexakis and Schlue in \cite{AS}. The purpose of this paper is to show that the result holds as well when gravity is coupled to electromagnetism. Alexakis and Schlue approach relies again on first proving stationarity to all orders at infinity as in \cite{BST}. Then, in order to extend this condition to the interior, they use unique continuation from infinity techniques based on Carleman estimates in the spirit of \cite{ASS}. For the first part of the proof, the exact null structure of the equations plays an important role in order to compute the metric, connection coefficients and curvature components to all orders at infinity with just the radiation field. For the second part a wave equation satisfied by the deformation tensor of the Weyl curvature, $\Lie_T C$, is derived and exploited to conclude stationarity in a neighbourhood of infinity by means of energy estimates. 	\\

We consider here the extended problem to include a Maxwell field. It is found that the results from \cite{AS} hold as well, that is, gravity and electromagnetism cannot balance each other to produce a periodic solution. The key point is that Maxwell equations also have a special non-linear structure and the coupled Einstein-Maxwell system can be treated in a similar way as before. The two deformation tensors, $\Lie_T C$ and $\Lie_T F$, also obey wave equations; however the coupling terms do not decay fast enough for the Alexakis-Schlue argument to work. Hence we are forced to revise and adapt their proof at the level of Carleman estimates to conclude the vanishing of the deformation tensors a little bit into the interior of the spacetime. The main result of this paper is, 

\begin{thm} \label{thm1}
Let $(\mathcal{M}, g, F)$ be an asymptotically flat (in the sense of definition \ref{def:AF}) non-radiating solution of the Einstein-Maxwell equations. Then there exists a time-like vector field  $T$ in a neighbourhood of spatial infinity such that 
	\[ \Lie_T g = 0 = \Lie_T F .	\]
\end{thm}	

We comment briefly about the assumptions. In this paper we will consider the class of asymptotically flat spacetimes admitting coordinates $(t, r, \theta^i)$ which are Minkowski to leading order and the decaying metric coefficients admit an infinite asymptotic expansion in terms of inverse powers of $r$, an area parameter. Moreover, it is required that these expansions are well-behaved with respect to derivatives up to second order (cf. definition \ref{def:AF}). We remark that this condition of asymptotic flatness is morally equivalent to smoothness at null infinity in the conformal picture, the advantage of this formulation lies in its adaptability to more general asymptotic conditions, e.g., polyhomogeneous expansions \cite{Winicour}. Indeed, the methods employed here can be generalised to cope with time-independent logarithmic singularities; and while we take into account the asymptotic behaviour obtained by Klainerman-Christodoulou up to order three, it is a definite hypothesis of Theorem \ref{thm1} that we work with metrics which are smooth at higher orders. For more details about the smoothness assumption we refer the reader to the introduction of Section \ref{chapter:asymptotics}. \\


By non-radiating we mean that the Bondi mass is constant in a neighbourhood of spatial infinity. Recall the Bondi mass loss formula for future null infinity:
	\[ \partial_u  M = - \frac{1}{32\pi} \int_{S^2} (|\Xi|^2 + |\underline{A}(F)|^2) d\mu_{\breve{\gamma}} , 	\]
where $\Xi$	and $\underline{A}(F)$ can be regarded as the gravitational and electromagnetic power radiated to infinity per unit solid angle, respectively. Moreover, they correspond to the leading order components of the Weyl and Faraday tensors, respectively. Hence, the condition of constant Bondi mass is equivalent to requiring the vanishing of the gravitational and electromagnetic radiation fields, i.e., the leading order terms. In fact as we will see in Lemma \ref{lem:decay non-radiating}, the components of the Weyl and Faraday tensors for non-radiating spacetimes are $\order(r^{-3})$ and $\order(r^{-2})$, respectively. We refer the reader to Section \ref{section:CK analysis} for more details. 	\\

The conclusion of Theorem \ref{thm1} can be seen from two point of views. Firstly, as a rigidity result for metrics with fast decaying curvature. Secondly, it is also a inheritance of symmetry result since the gravity and electromagnetic field both turned out to be time-independent. As we will see shortly, this is not the case for other matter-energy models coupled to gravity.	\\

We remark that the above result can be generalised to include a massless Klein-Gordon scalar field coupled with gravity and electromagnetism. A more precise statement accompanied with a sketched proof is presented in the Appendix. We stress that the fall-off conditions required for the whole system have to be assumed at this point due to the lack of a stability result for an Einstein-Klein-Gordon system. In Theorem \ref{prop:Klein-Gordon} we consider the fall-off obtained from the linear analysis performed in Winicour \cite{Winicour} together with a smoothness assumption at higher orders as in definition \ref{def:AF}.\\

Bi{\v{c}}{\'a}k-Scholtz-Tod also tackled this Einstein-Klein-Gordon problem in \cite{BST2} (a continuation of \cite{BST}). They again go around the fall-off condition by requiring analyticity of the fields all the way up to infinity and conclude that time-periodic asymptotically flat Einstein-massless-Klein-Gordon systems are in fact stationary. \\

The results presented in this paper are local around spatial infinity and hence are applicable to any system whose matter source is spatially compact. We mention here what is known about the case when the matter content extends to infinity. The simplest model to consider is that of a massive Klein-Gordon field coupled to Einstein equations. As mentioned above the techniques used in this work carry on to the Einstein-massless-Klein-Gordon system; however as we will see, the massive case is different.  \\

In \cite{Herdeiro14}, Herdeiro and Rady constructed time-periodic Einstein-Klein-Gordon solutions. They start by analysing linear perturbations of the Klein-Gordon equation on a fixed Kerr background and take them as starting points to solve the full non-linear equations numerically. The underlying family of spacetimes are asymptotically flat, stationary, axisymmetric and with regular horizons. The scalar field however does not share the symmetries since the ansatz is taken to be  $\phi(r,\theta)e^{i(m \varphi - \omega t)}$.  Thus, this spacetimes can be regarded as hairy black holes bifurcating off the Kerr solution. Later on, Chodosh and Shlapentokh-Rothman, \cite{Chodosh}, retook the problem and proved, analytically, the existence of such spacetimes in a small neighbourhood of the Kerr family. All this is in contrast with our result where the symmetry is inherited by all the fields. The crucial difference in the assumptions is the presence of a non-zero mass for the scalar field, which acts as a non-decaying potential. \\

Related to the above hairy black holes is the existence of countably many time-periodic, spherically symmetric asymptotically flat \textit{boson stars} given by Bizo\'n and Wasserman in \cite{Bizon}. These are solutions of the Einstein-Klein-Gordon equations where the underlying spacetime is static while the complex scalar field has a positive mass and takes the form of a standing wave $\phi(r)e^{i\omega t}$. Therefore the matter field does not inherit the time-like symmetry. \\
\\
\textit{Outline of the paper.} In \textbf{Section 2} we introduce null coordinates adapted to future null infinity. We do this following the work of Christodoulou-Klainerman \cite{CK}, Christodoulou \cite{Christodoulou91} and Klainerman-Nicol{\`o} \cite{KN}. In addition to providing a background, these coordinates also give us a time-like quasi-symmetry which will play the role of the candidate Killing field. \\

Also, a tetrad adapted to these coordinates is defined and the structure equations are written with respect to this basis. Then the gauge conditions are explicitly spell out and we briefly explain the null hierarchy of the equations. To end this section we summarise the  Christodoulou and Klainerman analysis based on the initial value formulation \cite{CK}. In particular their deduced asymptotic expansions and the concepts of mass and radiation are reviewed.	\\

In \textbf{Section 3} we use the tetrad formalism to achieve the first step in the proof of the main theorem. That is, it is proven that a non-radiating spacetime is stationary to all orders at infinity. We recursively compute the connection and curvature coefficients to all orders at infinity (here is where the smoothness assumption of asymptotic flatness in definition \ref{def:AF} plays an important role in the form of asymptotic expansions in inverse powers of $r$). The hierarchy found by BMS, also interpreted as signature levels in the language of Christodoulou-Klainerman, helps  to understand the limiting structure at future null infinity of the Einstein equations. Then the aforementioned hierarchy helps us identify levels where the equations become linear for the quantities belonging to that level. Moreover, the radiation fields can be regarded as the necessary initial data to run an induction argument and find recurrence relations. In particular, in the absence of radiation fields, all the asymptotic quantities are found to be time-independent. Also, the procedure sheds light on the way the different terms in the $r$-expansion of the metric and connection coefficients couple to each other via the Einstein equations. \\

The above results together with an analytical condition already imply the stationarity of non-radiating electrovacuum spacetimes in a neighbourhood of infinity as in \cite{BST}. However, one of the goals of this paper is to dispense with the analyticity assumption. We are able to remain in the smooth class by using the methods explained in Section 4 and conclude the stationarity of non-radiating spacetimes in the class of smooth metrics. \\

The technique to compute the asymptotic quantities to all orders at infinity is so basic that we feel compelled to give an account here of the main idea with a toy model: An out-going wave on Minkowski spacetime. The reader will find that the procedure is straightforward and the difficulty to apply it to the Einstein equations lies only in the intricacy of the equations themselves. \\
\\
\textit{Toy model.} Consider the Minkowski metric in spherical coordinates: 
	\[ \eta = -dt^2 + dr^2+ r^2 \breve{\gamma} , \]
where $\breve{\gamma}$ is the round metric on the sphere $\mathbb{S}^2$. We change to out-going null coordinates which are better suited for the problem at hand. That is, let $u=t-r$, then the metric can be written as
	\[ \eta = -du^2 - 2dudr + r^2 \breve{\gamma} . \]
Note that the level sets of $u$ are null hypersurfaces ruled by $\partial_r$. Now, let $\phi$ be a solution of the free wave equation, 
	$$\square_{\eta} \phi = 0. $$
Assume moreover that $\phi$ admits an expansion of the form\footnote{Such an expansion is morally equivalent to analyticity all the way to infinity; an assumption arguably incompatible with the wave equation. Here we assume it for the sake of simplicity. The role of the wave equation in this argument will be merely computational as opposed to evolutionary. An asymptotic expansion would work as well.}
	\[ \phi (u,r,\theta^2, \theta^3) = \sum_{n=1}^{\infty} \overset{(n)}{\phi} (u, \theta^2, \theta^3) \frac{1}{r^n} ,\]
which is well-behaved with respect to derivatives.\\

We wish to find $\overset{(n)}{\phi}$ in terms of $\overset{(1)}{\phi}$ and $\overset{(n)}{\phi_0} := \lim_{u\rightarrow -\infty} \overset{(n)}{\phi}$. To do so we consider $X = \grad \phi$ and rewrite the wave equation as the 1st order system, 
	\begin{align*}
		\curl X = 0 , \\
		\divergence X = 0 .
	\end{align*}
For easy comparison with the CK notation we define the null components of $X$ as
	\[ x:= X_r, \quad \underline{x}:=X_u, \quad \cancel{X} := \frac1r \breve{\grad} \, \phi	\]
For the time being let us denote by $\breve{\grad}$, $\breve{\divergence}$ and $\breve{\Delta}$  the gradient, divergence and Laplace operators on the unit round sphere, respectively. In coordinates, the previous equations read
	\begin{align*}
		\partial_r \cancel{X} &= \frac1r \breve{\grad} \, x -\frac1r \cancel{X} ,	\quad	&	\partial_u \cancel{X} &= \frac1r \breve{\grad} \, \underline{x} -\frac1r \cancel{X}	, \\
		\partial_r \underline{x} &= \partial_u x, \quad	&	\breve{\curl} \, \cancel{X}  &= 0	, 					\\
		\partial_u x &= - \partial_r \underline{x}  + \frac{1}{r^2}\partial_r(r^2 x) + \frac1r \breve{\divergence} \, X .
	\end{align*}
These equations (the first column) imply the following recurrence relations for the asymptotic quantities, 
	\begin{align*}
		-(n-1) \overset{(n)}{\cancel{X}} &= \breve{\grad} \overset{(n)}{x}, 	\\
		-n \overset{(n)}{\underline{x}} &= \partial_u \overset{(n+1)}{x}, 	\\
		 \partial_u \overset{(n+1)}{x} &= n\overset{(n)}{\underline{x}} -(n-2) \overset{(n)}{x} -\breve{\divergence} \, \overset{(n)}{\cancel{X}}.
	\end{align*}
From these we can deduce an evolution equation for each $\overset{(n+1)}{x}$ in terms of lower order data, 
	\begin{align*}
	2\partial_u \overset{(n+1)}{x} = \frac{1}{n-1} \breve{\Delta}  \overset{(n)}{x}  - (n-2)  \overset{(n)}{x} ,
	\end{align*}
which can be solved given initial data $\overset{(n+1)}{x_0}$ at $u= -\infty$. Note that the asymptotic expansion for $\phi$ implies $ \overset{(1)}{x}=0$ and $ \overset{(2)}{x}=-\overset{(1)}{\phi}$, in particular, we can compute $\overset{(n+1)}{x}$ inductively for $n>2$ assuming we know the radiation field $\overset{(1)}{\phi}$. \\

Therefore, we can compute $ \overset{(n)}{x}$, $ \overset{(n)}{\underline{x}}$ and $ \overset{(n)}{\cancel{X}}$ for all $n\in \N$ along future null infinity in terms of the radiation field $\overset{(1)}{\phi}$ and the ``pole moments at spatial infinity'' $ \overset{(n)}{x_0}$, $n\in \N$.   $_{\blacksquare}$	\\

This toy model reflects perfectly well the asymptotic relations when the equations can be extended smoothly up to infinity. The analysis for Maxwell and Einstein equations is completely analogous albeit more technical. The previous relations will be used in the Appendix when we discuss the extension of the main theorem to include a massless Klein-Gordon scalar field.	\\

In \textbf{Section 4} we complete the proof of the main theorem. We explain how the previous asymptotic values at infinity can be translated to a local result in the spacetime. That is, the condition of stationarity to all orders at infinity is extended to a neighbourhood of spatial infinity. \\

The main technical tool is that of Carleman estimates. These are a priori wave estimates for functions with fast decay (faster than any polynomial) towards infinity. They can also be thought as energy estimates adapted to time-like boundaries conditions where now we aim at controlling the bulk term while making the boundary terms vanish. A key ingredient in these inequalities is a 1-parameter weight function which accounts for the exponential rate of decay. With the help of this weight, it is possible to bound the $L^2$-norms of a function $\phi$ and its first derivatives in terms of $\square \phi$, provided that the boundary of the region under consideration satisfies the so-called pseudo-convex condition. \\

At this point we make a parenthesis to look for wave equations satisfied by the components of the deformation tensors $\Lie_T C$ and $\Lie_T F$. These are the Ionescu-Klainerman tensorial equations, \cite{IK}. We need to revise them in order to cope with a non-vacuum spacetime. These wave equations, written in suitable coordinates,  in conjunction with the Carleman estimates are used to find $L^2$-bounds for the components of the deformation tensors and its first derivatives. These bounds depend on an arbitrarily large parameter which can be taken to infinity to conclude the vanishing of the functions in the interior. \\

Finally, an \textbf{Appendix} is included where we state a variation of Theorem \ref{thm1} to include a Klein-Gordon field, as well as the corresponding recurrence relations. We also include a brief review of the tetrad formalism employed in this paper, as well as the CK notation. Finally a table is included to compare the CK and NP notations.

\section{Coordinates and gauge conditions} \label{section:tetrad and coordinates}
\setcounter{footnote}{1}


Here we construct the necessary out-going null coordinates in the context of asymptotically flat spacetimes. These are given by a foliation by null hypersurfaces complemented with a transversal coordinate given by the candidate Killing field, which is roughly speaking, a time-like symmetry to first order. The existence and construction of such foliation is taken from Christodoulou \cite{Christodoulou91}. \\

Let $(\Sigma_0, h,K)$ be a strongly asymptotically flat initial data set and consider an exhaustion of $\Sigma_0$ by balls $B_d$, where $d$ is an asymptotically flat radial coordinate on $\Sigma_0$. Within the corresponding Cauchy development define $C_d^-$, resp. $C_d^+$, to be the ingoing, resp. outgoing, null hypersurface consisting of geodesic rays emanating from $S_d =\partial B_d$. Fix some $S_0=\partial B_{d_0}$, thanks to Christodoulou-Klainerman \cite{CK}, we can choose it so that the generators of $C_0^{\pm} := C_{d_0}^{\pm}$ have no future end points. Finally let $S_0^* = C_0^+ \cap C_{d^*}^-$ for some $d^*$ large.	\\

At $S_0^*$ we can define the future-directed null outgoing and ingoing normal vectors $L$ and $\underline{L}$, respectively, by the conditions:
	\[	\langle L, \underline{L}\rangle =-2 , \quad \langle L,L \rangle = \langle \underline{L}, \underline{L}\rangle = \langle L, X \rangle = \langle \underline{L}, X \rangle = 0, \quad X\in \mathcal{T}_pS_0^* . 	\]
These conditions fix $L, \underline{L}$ up to rescaling by a function $a$ on $S_0^*$:
	\[ L \mapsto a L , \qquad \underline{L} \mapsto \frac1a \underline{L} .	\]

Next, $(L, \underline{L})$ are extended to $C_{d^*}^-$ as follows: Let $\tilde{\underline{L}}$ be the extension of $\underline{L}$ by geodesic flow, that is, 
	\[ \nabla_{\tilde{\underline{L}}}\tilde{\underline{L}} = 0, \quad \textrm{along } C_{d^*}^- \text{, with} \quad \tilde{\underline{L}} = \underline{L} \quad \textrm{on } S_0^* .\]
With the help of its affine function $t$ (that is, $\tilde{\underline{L}} \cdot t = 1$ on $C_{d^*}^-$ and $t=0$ on $S_0^*$) we define the \emph{retarded time} function $u$ on $C_{d^*}^-$ to be the area parameter on a level set of $t$,
	\[	u(t) := 2 (r^*_0 - r^*_t) ,	\]
here $r^*_t$ is the area radius of $S_t^*$, the level set of $t$ on $C_{d^*}^-$. Now, we set $\underline{L}$ to be the velocity field of $u$	on $C_{d^*}^-$, that is, $\underline{L} = \frac{dt}{du}\tilde{\underline{L}}$ along the null geodesics ruling $C_{d^*}^-$. In particular, $\underline{L}(u)=1$.\\

Then, let $L$ be defined on $C_{d^*}^-$ as the null conjugate to $\underline{L}$ with respect to level sets of $u$, $S_u^*$:
	\[ \langle L, \underline{L}\rangle = -2, \quad \langle L, X \rangle = 0, \quad X\in \mathcal{T}_pS_u^* . 	\]
Finally we extend $u$ by choosing a solution of the eikonal equation,
	\[ g^{\al \be} \partial_{\al} u \partial_{\be} u = 0	\]
with the already prescribed value of $u$ along $C_{d^*}^-$. Let $C_u^+$ be the level sets of $u$, we extend $L$ by geodesics,
	\[ \nabla_L L = 0 , \quad \textrm{on the relevant open set.}\]
Let $s$ be the affine parameter: $L\cdot s = 1$ and $s|_{S_u^*}=r_u^*$.	\\

The above procedure also gives coordinates $(s,u, \theta^i)$, by choosing $(\theta^i)$ on $S_0^*$ then $u-$flowing them along $\underline{L}$ on $C_{d^*}^-$ to obtain coordinates $(u, \theta^i)$ and then $s-$flowing the latter along $L$ to cover an open set near infinity. \\

It worth noticing that the above construction is engineered so that $L$, $\underline{L}$ and the candidate Killing field $\partial_u$ are related to leading order in the usual way, that is, 
	\begin{align*}
		\partial_u = \frac12 \left( L + \underline{L}\right) + \order(r^{-1}).
	\end{align*}

    \begin{figure}[h]
        \begin{center}
\begin{tikzpicture}[scale=.8]		
	\draw (-3,0) node {};
	\fill (4,0) circle (2pt) node[right] {$i^0$};

	\draw [line width = 1pt, dashed] (0,4) -- (4,0) node[pos=.5, above right]{$I^+$};
	\draw [line width = 1pt, dashed] (0,-4) -- (4,0) node[pos=.5, below right]{$I^-$};

	\draw [line width = 1.5pt, ] (-2,0) -- (4,0) node[pos=0, left]{$\Sigma_0$};
	
	\draw (1,2) circle (3pt) node[left] {$S_0^*$};
	\draw [line width = 1pt] (-1,0) -- (1.5,2.5) node[pos=0.3, left ]{$C_0^+$};
	\draw [line width = 1.5pt] (0,3) -- (3,0) node[pos=0.8, left]{$C_{d^*}^-$};

	\draw[->] [line width = 1pt] (7,1) -- (6,2) node[pos=1, above]{$\underline{L}$};
	\draw[->] [line width = 1pt] (7,1) -- (8,2) node[pos=1, above]{$L$};

	\draw [line width = 1pt, dashed] (7,-2) -- (6,-1) ;
	\draw[->] [line width = 1pt] (7,-2) -- (6.7,-.8) node[pos=1, above]{$u$};
	\draw[->] [line width = 1pt] (7,-2) -- (8,-1) node[pos=1, above]{$s$};
		
\end{tikzpicture}  
    \caption{Coordinates in a neighbourhood of future null infinity. The level sets of $u$ are the outgoing null hypersurfaces $C_u^+$ ruled by $L$. The level sets of $s$ are timelike.}
\end{center}
	\end{figure}
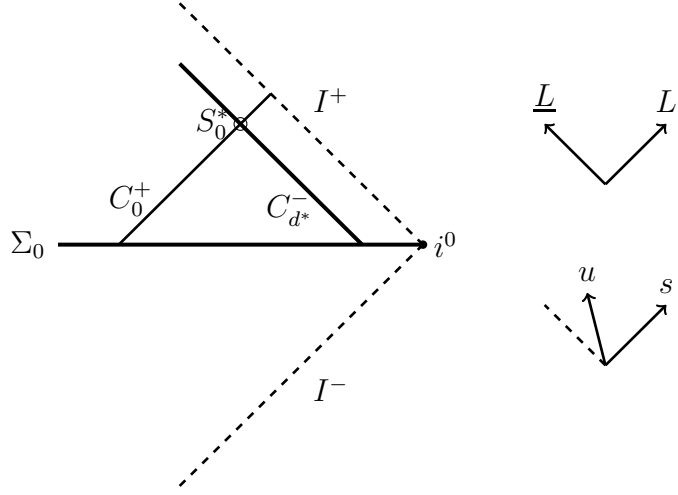
	
\subsection{Gauge Conditions}

It will be convenient to work with a tetrad
	\begin{align} \label{tetrad}
		\begin{aligned}
		e_0 & := L = \partial_s, \\
		e_1 & := \underline{L} = f^0\partial_s + f^1 \partial_u + f^i\partial_{\theta^i}, \\
		e_i & := {h_i}^j \partial_{\theta^j}	,
		\end{aligned}
	\end{align}
satisfying
			\begin{eqnarray} \label{metric-matrix}
	 \eta_{\mu\nu} := \langle e_{\mu}, e_{\nu} \rangle =  \left( \begin{array}{ccc}
	0 & -2 &  0 \\
	-2 & 0 &  0	\\
	 0 & 0 & \textrm{Id}_{2}\\
		 \end{array}\right) .
		\end{eqnarray}
The above change of basis can be written compactly in terms of an \emph{orthonormalisation matrix}, 
	\begin{align*}
		e_{\mu} = {h_{\mu}}^{a} \partial_{x^{a}} .
	\end{align*}
That is, 
	\begin{eqnarray*} 
	 {h_{\mu}}^{a} := \left(  \begin{array}{cc}
	 	\begin{array}{cc}  	1 & f^0  \\  0 & f^1  	\end{array}  & 
	 	\begin{array}{cc}  	0 & 0  \\	0 & 0  	\end{array} 	 \\ 
	 	\begin{array}{cc}  	0 & f^2  \\  0 & f^3  	\end{array}  & 
	 	\begin{array}{c}  	 {h_{i}}^k 	\end{array}  
	 	\end{array} \right) .
	\end{eqnarray*}
The metric components are related to the orthonormalisation matrix by $g^{ab} = {h_{\mu}}^{a} \eta^{\mu \nu} {h_{\nu}}^{b}$, where $g^{ab}$ is the inverse matrix of $g_{ab} := \langle \partial_{x^{a}}, \partial_{x^{b}} \rangle$. Explicitly, 
	\begin{eqnarray*} 
	 g^{ab}  = \left(  \begin{array}{cc}
	 	\begin{array}{cc}  	0 & -\frac12 f^1  \\  -\frac12 f^1 & -f^1 f^0 + (f^2)^2 + (f^3)^2  	\end{array}  & 
	 	\begin{array}{cc}  	0 & 0  \\	f^i {h_i}^2 &  f^i {h_i}^3  	\end{array} 	 \\ 
	 	\begin{array}{cc}  	0 \qquad \qquad \; &  f^i {h_i}^2 \qquad \quad \\  0 \qquad \qquad \; &  f^i {h_i}^3 \qquad \quad  	\end{array}  & 
	 	\begin{array}{c}  	 {h_{i}}^k \delta^{ij} {h_j}^l 	\end{array}  
	 	\end{array} \right) .
	\end{eqnarray*}

\textbf{Remark.} The function $f^1$ is constant and we can choose it so that $f^1=2$. This can be showed by first noting that $\partial_s \langle \partial_u, \partial_s \rangle = \langle \nabla_{\partial_s} \partial_u, \partial_s \rangle + \langle \partial_u, \nabla_{\partial_s} \partial_s \rangle = \langle \nabla_{\partial_u} \partial_s,  \partial_s \rangle  = \frac12 \partial_u \langle \partial_s, \partial_s \rangle = 0$, hence $\langle \partial_u, \partial_s \rangle$ is constant along $s-$lines. Then the assertion follows from $-2 =\langle L, \underline{L} \rangle =  f^1 \langle \partial_u, \partial_s \rangle$. \\

We call all these choices the \emph{null-geodesic gauge conditions}. \\
	
The connection coefficients are the components of the derivative operator with respect to the tetrad $e_{\mu}$:
	\begin{eqnarray*}
		\omega_{\lambda \mu \nu} := \langle \nabla_{e_{\lambda}} e_{\nu}, e_{\mu}	\rangle .
	\end{eqnarray*}
They satisfy $\omega_{\lambda \mu \nu} = -\omega_{\lambda \nu \mu}$, this is a consequence of $\eta_{\mu \nu}$ being a constant matrix. Moreover, the connection coefficients and the orthonormalisation matrix are related by the \emph{frame equations} (the torsion-free property of the connection in this language, see Appendix for details). In the null gauge case they read
	\begin{align}
		\omega_{001} &= 0, 					\quad	&	-2\zeta_i = \omega_{10i} &= \omega_{i01} = -V_i, \\
		\omega_{00i} &= 0 	,				\quad	&	 -\chi_{ij}= \omega_{i0j} &= \omega_{j0i} = -\chi_{ji} , \\
		-2\underline{\zeta}_i=\omega_{01i} &= \omega_{i10}=V_i	,	\quad	&	- \underline{\chi}_{ij} = \omega_{i1j} &= \omega_{j1i} = - \underline{\chi}_{ji}, 	\\
		e_0(f^0) &= \frac12 \omega_{110} = 2 \underline{\omega},	\quad	&	e_i(f^0) &= \frac12 \omega_{11i} =\la_i, \\
		e_0(f^i) &= ( {\omega_0}^j{}_1 - {\omega_1}^j{}_0 ){h_j}^i	,\quad	&	e_i({h_j}^l) - e_j({h_i}^l) &= ({\omega_i}^k{}_j - {\omega_j}^k{}_i ){h_k}^l, 	\\
		e_0({h_i}^j) &= ( {\omega_0}^l{}_i - {\omega_i}^l{}_0 ){h_l}^j,	\quad	&	e_1({h_i}^j) - e_i(f^j) &= ( {\omega_1}^l{}_i - {\omega_i}^l{}_i ){h_l}^j. 
	\end{align}
We refer the reader to the Appendix for a review of the CK notation and a comparison with the NP spin coefficients and the slight variations adopted in this paper.	\\
	
The previous construction of coordinates and tetrad fixes $e_0=L$ and $e_1=\underline{L}$ up to rescaling
	\[ L \mapsto a L , \qquad \underline{L} \mapsto \frac1a \underline{L} ,	\]
where $a$ is a function of the sphere $S^*_0$. In view of the corresponding scaling of the second fundamental forms,
	\[ \chi \mapsto a \chi , \qquad \underline{\chi} \mapsto \frac1a \underline{\chi} .	\]
The function $a$ can be chosen so that
	\[ \lim_{r \rightarrow \infty} ( \textrm{tr} \chi + \textrm{tr} \underline{\chi} ) = 0	. \]
In other words, it is required that the first variation of area of $S_0^*$ along geodesics defined by the direction of $L+\underline{L}$ tends to $0$ at infinity. We define then
	\[T := \partial_u	\]
to be the candidate Killing vector field. \\


Finally, we can also arrange for $\omega_{0ij}=0$ as in Lemma \ref{lem:comoving gauge}. This is done by solving an ODE, so we still have the freedom to choose an initial condition for the system. This translates to a freedom to adapt the frame $\{e_i\}$ at infinity. More precisely, we are still free to choose the leading coefficient in the asymptotic expansion of ${h_i}^j$. Since we are approaching the round unit-sphere at infinity, we can set
	\begin{eqnarray*} 
	 \overset{(1)}{{h_i}^j} = \left( \begin{array}{cc}
	 1 & 0  \\
 	 0 & \frac{1}{\sin\theta} \\
	 \end{array}\right) .
	\end{eqnarray*}
We call the choice $\omega_{0ij}=0$ the \emph{co-moving frame gauge}. 	\\

In the next section we work with the above choice of gauge conditions. While the null-geodesic gauge is essential for the procedure, the co-moving frame gauge is used mostly for convenience. The arguments go through provided that we think of $\omega_{0ij}$ and $ \overset{(1)}{{h_i}^j}$ as given data.	\\

Note that the frame equations imply that $\chi_{ij}, \underline{\chi}_{ij}$ are symmetric 2-tensors, $2\zeta_i=V_i=-2\underline{\zeta}_i$, $\xi_i=0$ and $\omega=0$. The evolution of the orthonormalisation matrix is then given by 
	\begin{align}
		e_0(f^0) &= 2 \underline{\omega},	\quad	&	e_i(f^0) &= \la_i, \\
		e_0(f^i) &= -\zeta^k {h_k}^i	,\quad	&	e_i({h_j}^l) - e_j({h_i}^l) &= ({\omega_i}^k{}_j - {\omega_j}^k{}_i ){h_k}^l, 	\\
		e_0({h_i}^j) &= -{\chi_i}^k{h_k}^j,	\quad	&	e_1({h_i}^j) - e_i(f^j) &= ( {\omega_1}^l{}_i - {\omega_i}^l{}_i ){h_l}^j. 
	\end{align}

\subsection{Structure equations}

Recall the formula for the Riemann curvature tensor components in terms of the connection coefficients:
	\[ R_{\rho \sigma \mu \nu} = e_{\rho} (\omega_{\sigma \mu \nu} ) -e_{\sigma} (\omega_{\rho \mu \nu} ) - {\omega_{\rho \mu}}^{\alpha} \omega_{\sigma \nu \alpha} + {\omega_{\sigma \mu}}^{\alpha} \omega_{\rho \nu \alpha}  - {\omega_{\sigma \rho}}^{\alpha} \omega_{\alpha \mu \nu} + {\omega_{\rho \sigma}}^{\alpha} \omega_{\alpha \mu \nu}  .  \]

The splitting into trace and traceless part is the usual
	\[ R_{\rho \sigma \mu \nu} = C_{\rho \sigma \mu \nu} + \frac{1}{2} ( \eta_{\rho \mu} S_{\sigma \nu} - \eta_{\sigma \mu} S_{\rho \nu} + \eta_{\sigma \nu} S_{\rho \mu} - \eta_{\rho \nu} S_{\sigma \mu} ) ,  \]
where $C_{\rho \sigma \mu \nu}$ and $S_{\mu \nu}$ are the components of the Weyl and Schouten tensors, respectively.  \\

The \emph{structure equations} are the combination of these two preceding equations, we write them schematically as
	\begin{align}	\label{structure equations}
		C + \eta \wedge S = e \wedge \omega + \omega \wedge \omega .	
	\end{align}
The Einstein equations appear implicitly when we regard the Schouten tensor as the stress-energy tensor of the matter/energy content. \\

In order to understand the structure equations in the null-geodesic gauge we introduce the concept of the signature of a tensor, as discussed in \cite{CK} p 148. We just quote here their clear exposition:	\\

\emph{Given any covariant tensor $U$ at a point of spacetime, we define a null component of it to be any tensor tangent to the sphere $S_{u,s}$ at a point, which is derived from $U$ by contractions with either $e_0$ or $e_1$ and projections to $S_{u,s}$. To any such component we assign a signature that it is defined as the difference between the total number of contractions with $e_0$ and the total number of contractions with $e_3$. We are now ready to state the following heuristic principle.	\\
\\
\textbf{Principle of Conservation of signature:} Consider an arbitrary covariant tensor $U$ that can be expressed as a multilinear form in an arbitrary number of covariant tensors $U_1 \ldots U_p$, with coefficients depending only on the spacetime metric and its volume form. Then the signature of any null term of U, expressed in terms of the null components of $U_1\ldots U_p$, is equal to the sum of the signatures of each constituent in the decomposition.}\\

In particular the signature of a tensor does not change by lowering/raising indices. We have included in Table 1 of the Appendix the signature of the different null components relevant to us. The above principle will allow us to guess the quadratic terms appearing in the structure equations without performing the calculations. Indeed, we already know that in the null-geodesic gauge the only connection coefficient with signature equal 2 vanishes; this fact is the one responsible for the success of the hierarchy-induction argument explained in Section \ref{section:quantities} to obtain the asymptotic quantities to all orders at infinity.\\

Now we illustrate how to apply this principle to the structure equations. We will denote by $U(l)$ a null component of $U$ with signature $l$. The signature 2 case yields schematically:
	\begin{align*}
		C(2) + \eta \wedge S(2) &= e(1) \omega(1) + e(0)\omega(2) + \omega(1) \wedge \omega(1) + \omega(0) \wedge  \omega(2), \\
				&=e(1) \omega(1) + \omega(1) \wedge \omega(1) . 
	\end{align*}			
where in the second line we have used the fact that the only connection coefficient with signature 2, $\omega_{0\al0}$, vanishes in the null-geodesic gauge. More generally we have the following: \\
\\
\emph{\textbf{Hierarchy of structure equations:} When working in the null-geodesic gauge there is no connection coefficient of signature $l-2$ on the right-hand side of the structure equation of signature $l$, $l=2,1,0$.}\\

This is precisely the hierarchy found by BMS, \cite{BMS}. The exact form of the structure equations is presented in equations (\ref{hierarchy0})-(\ref{hierarchy4}).

\subsection{Christodoulou-Klainerman estimates } \label{section:CK analysis}



In their book The Global Nonlinear Stability of the Minkowski Space \cite{CK}, Christodoulou and Klainerman proved that any strongly asymptotically flat initial data set that satisfies a global smallness assumption, leads to a unique globally hyperbolic and geodesically complete solution of the Einstein vacuum equations. Moreover the development is globally asymptotically flat in the sense that the curvature vanishes at infinity in all directions. \\

Here we list some conclusions of their work regarding the asymptotic behaviour of solutions suitably close to Minkowski spacetime. Of particular interest are the decay rates obtained for the connection coefficients and the null components of the Weyl tensor. It is observed that the peeling property may not hold in general. \\

While all the previous was done in the vacuum case, the extension to include an Electromagnetic field was worked out by Zipser in her PhD dissertation \cite{Zipser}. The same conclusion for an Einstein-massless-Klein-Gordon system is expected to hold; however to the author's knowledge such result is still lacking in the literature. \\

Using the null out-going coordinates as constructed in the previous section, Christodoulou-Klainerman showed that, for fixed $u$, $S_{s,u}$ converges to the standard round sphere embedded in Minkowski, that is, let $\cancel{g}$ be the induced metric on $S_{s,u}$, then
	\begin{align*}
	\lim_{u; r \rightarrow \infty} r^{-2} \cancel{g} &= \breve{\gamma}, \qquad &	\lim_{u; r \rightarrow \infty} K[r^{-2} \cancel{g} ] &= 1, \\
	\lim_{u; r \rightarrow \infty} r \tr \chi &= 2, \qquad &	\lim_{u; r \rightarrow \infty} r \tr \underline{\chi} &=-2 , 
	\end{align*}
here the limit $\lim_{u; r \rightarrow \infty}$ is taken along a level set of $u$ while letting the area function $r$ tend to infinity; $\breve{\gamma}$ is the standard round metric on $\mathbb{S}^2$ and $K$ is the Gauss curvature.\\

Denote by $\hat{\chi}$ the trace-free part of $\chi$, then to next order they obtained
	\begin{align*}
	\lim_{u; r \rightarrow \infty} r^{2} \hat{\chi} &= \Sigma, \qquad &&	\lim_{u; r \rightarrow \infty} r(r\tr\chi - 2) = H, \\
	\lim_{u; r \rightarrow \infty} r \underline{\hat{\chi}} &= \Xi, \qquad &&	\lim_{u; r \rightarrow \infty} r(r\tr\underline{\chi} - 2) = \underline{H} ,
	\end{align*}
where $\Sigma$, $\Xi$ are symmetric trace-less $u$-dependent 2-covariant tensors defined on $\mathbb{S}^2$ and $H$, $\underline{H}$ are $u$-independent functions on $\mathbb{S}^2$ of vanishing mean. Moreover, 
	\begin{align*}
		|\Xi(u, \cdot)|_{\breve{\gamma}} \leq C(1 + |u|)^{-3/2}.
	\end{align*}
Recall the null decomposition the Weyl tensor with respect to a tetrad $\{e_0, e_1, e_i \}$,
	\begin{align} \label{null Weyl}
		\alpha_{ij} &= C_{i0j0} ,   \qquad	&&	\underline{\alpha}_{ij} = C_{i1j1} ,  \\
		2 \beta_i &= C_{i010} ,   \qquad	&&	2 \underline{\beta}_i = C_{i110} ,  \\
		4 \rho &= C_{1010} , 	\qquad	&&	\sigma = C_{0123} .
	\end{align}
The curvature null components also decay. Explicitly, there exist a symmetric trace-less 2-tensor $\underline{A}$, a 1-form $\underline{B}$ and functions $P$, $Q$ defined on $\mathbb{S}^2$ and $u$-dependent satisfying
	\begin{align} \label{null Weyl2}
	\lim_{u; r \rightarrow \infty} r \underline{\alpha} &= \underline{A}, \qquad &	\lim_{u; r \rightarrow \infty} r^2 \underline{\beta} &= \underline{B}, \\
	\lim_{u; r \rightarrow \infty} r^{3} \rho &= P , \qquad &	\lim_{u; r \rightarrow \infty} r^3 \sigma &= Q . 
	\end{align}
Moreover, these limits decay in $u$ as follows,
	\begin{align}
		|\underline{A}(u, \cdot)| & \leq C (1+|u|)^{-5/2} , \qquad	&	|\underline{B} (u, \cdot)| &\leq C(1+|u|)^{-3/2}, 		\\
		|P(u, \cdot) - \overline{P}(u)| & \leq (1+|u|)^{-1/2} , \qquad	&	|Q(u, \cdot) - \overline{Q}(u)| &\leq (1+|u|)^{-1/2}, 
	\end{align}
and $\lim_{u \rightarrow - \infty} \overline{P}(u)=0$, $\lim_{u \rightarrow - \infty} \overline{Q}(u)=0$.\\

This is all consistent with the presence of peeling. In contrast, they could only deduce a weaker fall-off for the remaining curvature components, namely, 
	\begin{align*}
		|\alpha|_{\cancel{g}} = \mathcal{O} (r^{-7/2}), \qquad		|\beta|_{\cancel{g}} = \mathcal{O} (r^{-7/2}),
	\end{align*}
For completeness we mention that the presence of the full peeling property would be achieved by $|\alpha|_{\cancel{g}} = \mathcal{O} (r^{-5})$ and $|\beta|_{\cancel{g}} = \mathcal{O} (r^{-4})$. \\

Finally, the following relations are implied by the structure equations,
	\begin{align*}
		\partial_u \Xi &= -\frac12 \underline{A}, 	\qquad	&	\breve{ \cancel{\divergence} }\ \Xi &=  \underline{B}, 	\\
		\partial_u \Sigma &= -\frac12 \Xi, 		\qquad	& 		\breve{ \cancel{\divergence} }\ \Sigma &=  \frac12 \breve{\cancel{\nabla}} H + Z, 	\\
		\partial_u \underline{H} &= -\frac12|\Xi|^2 ,
	\end{align*}
	where $\breve{\cancel{\nabla}}$ and $\breve{\cancel{\divergence}}$ stand for the connection and divergence operators on the round sphere, respectively.
	
\subsubsection{Mass}	

The \emph{Hawking mass} enclosed by a 2-sphere $S_{s,u}$ is defined to be, 
	\begin{align*}
		m(s,u) = \frac{r(s,u)}{2} \left( 1 + \frac{1}{16\pi} \int_{S_{s,u}} \tr \chi \tr \underline{\chi}	\right),
	\end{align*}
Christodoulou and Klainerman also deduced from their analysis the decay rate of the Hawking mass as well as its evolution equation. Explicitly, there exists a function $M(u)$ such that along a level set of $u$ we have
	\begin{align*}
		m(s,u) = M(u) + \order (r^{-1}).
	\end{align*}
The function $M(u)$ is called the Bondi mass. Furthermore, it satisfies the Bondi mass formula
	\begin{align*}
		\partial_u M = - \frac{1}{32\pi} \int_{\mathbb{S}^2} |\Xi(u, \cdot)|^2 d\mu_{\breve{\gamma}} . 
	\end{align*}

\subsubsection{Gravity coupled with electromagnetism}	

The previous analysis was done for vacuum solutions suitably close to Minkowski spacetime. The generalisation to an Einstein-Maxwell system was carried out by Zipser in her PhD thesis \cite{Zipser}. She found that the null components of the Faraday tensor decay as expected from the linear analysis (cf. \cite{CKlinear}).	\\

Recall the null decomposition of the Faraday tensor.
\begin{align} \label{null Faraday}
	\alpha(F)_i &= F_{i0} ,			\quad& 	\underline{\alpha}(F)_i &= F_{i1} ,	\\
	\rho (F)	&= \frac12F_{10},	\quad&		\sigma(F) &= F_{23} .
\end{align}

As a consequence of Zipser's analysis we have that the following limits exist,
	\begin{align} \label{null Faraday2}
	\lim_{u; r \rightarrow \infty} r \underline{\alpha}(F) &= \underline{A}(F), \\
	\lim_{u; r \rightarrow \infty} r^{2} \rho(F) &= P(F) ,	\\
	\lim_{u; r \rightarrow \infty} r^2 \sigma (F) &= Q(F) . 
	\end{align}

However, full peeling is not available for the remaining component. Instead we have the weaker estimate along a level set of $u$, 
	\begin{align}  \label{null Faraday3}
	\alpha(F) = \order(r^{-\frac52}) .
	\end{align}

The generalisation made by Zipser has as a consequence that the Hawking mass also has a limit, that we may call Bondi mass, and this limit satisfies the following Bondi mass formula:
	\begin{align} \label{Bondi mass formula}
		\partial_u M = - \frac{1}{32\pi} \int_{\mathbb{S}^2} |\Xi(u, \cdot)|^2 + |\underline{A}(F)(u, \cdot)|^2 d\mu_{\breve{\gamma}} .	
	\end{align}

For this reason we call $\Xi_{ij}$ and $\underline{A}(F)_i$ the \emph{radiation fields}. 

\begin{definition} \label{non-radiating}
	\emph{Non-radiating spacetimes.} An asymptotically flat solution to the Einstein-Maxwell equations is called non-radiating (towards future null infinity) if the Bondi mass $M(u)$ is constant as a function of retarded time $u$. 
\end{definition}

This is strongly related to the order of decay of the curvature at infinity. Indeed we have the following lemma which is a consequence of Corollary \ref{cor:non-radiating}.

\begin{lem} \label{lem:decay non-radiating}
Assume we are working with the tetrad, coordinates and gauge conditions imposed in this section, then an asymptotically flat solution of the Einstein-Maxwell equations is non-radiating towards future infinity if and only if
	\begin{align*}
	C_{\rho \sigma \mu \nu} = \order(r^{-3})  \quad \textrm{and} \quad	F_{\mu \nu} = \order(r^{-2}) ,
 	\end{align*}
\end{lem}

\section{Asymptotic behaviour of the fields} \label{chapter:asymptotics}
\setcounter{footnote}{1}

In this section we start by presenting the main technical assumptions required to `push' the Einstein equations to infinity at all orders. Namely we estate the class of functional spaces and asymptotically flat spacetimes that we will consider. Without further comment all the spacetimes in this paper will be assumed asymptotically flat in the sense of definition \ref{def:AF} below. \\

Next, we compute asymptotic quantities at infinity. We follow  Bi\v{c}\'ak-Sholtz-Tod \cite{BST} and Alexakis-Schlue \cite{AS} in their asymptotic analysis and extent their results to show that the radiation fields determine the evolution of the metric to all orders at infinity. Indeed, by assuming an asymptotic expansion we can take the limit for each order of the structure equations and find algebraic and $u$-transport relations at infinity at a given order. These relations are well-suited for an induction process provided that we know the radiation fields, $\Xi_{ij}$ and $\underline{A}(F)_i$ along null infinity and initial conditions for $\overset{(2)}{\rho} $ and $\overset{(3)}{\zeta_i} $ at a sphere. The main results of this section state that these data determines the solution to all orders at infinity, see Proposition \ref{prop:recurrence relations} and its ensuing corollaries.	\\

Before giving our definition of asymptotic flatness, we digress briefly about the different coordinate systems that we will be using. These are time-space, null out-going and double-null coordinates. 
	\begin{itemize}
	\item \emph{Time-space coordinates.} These are important as they split naturally the initial data and the evolution equations. It is a definite hypothesis of this work that the asymptotic properties at spatial infinity of the Cauchy data are preserved under the evolution; this is the content of part $i)$ in definition \ref{def:AF}, below. It would be desirable to deduced this fact\footnote{That is, that smoothness at spatial infinity is preserved for finite but arbitrarily large time. Then, the desired expansion would hold for, e.g., time-periodic spacetimes.} from the Einstein equations; this would require a `preservation of regularity estimate for all orders at spatial infinity' which appears to be missing in the literature. \\
	
	We stress here that the CK analysis is not suited for this problem as their gauge conditions are naturally adapted to null inifnity. On the other hand, Friedrich's analysis of his conformal equations, \cite{Friedrich02}, provides strong evidence about the preservation of regularity at spatial infinity, since the equations can be extended regularly there.\\ 
	
	
	\item \emph{Null out-going coordinates.} Their construction is one of the crucial steps in the CK analysis. These coordinates are used to define the 15-parameter family of quasi-symmetries (corresponding to translations, Lorentz rotations, scaling and inverted translations in Minkowski spacetime) which serve to compare all geometric quantities to that of flat spacetime. Of particular importance to us is the construction of the time-like quasi-translations, which will play the role of candidate Killing field.\\
	
	In \cite{CK} they define $u$ by setting initial conditions on a ``last Cauchy hypersurface'' and then extending by geodesics. Here we have followed Christodoulou \cite{Christodoulou91} and Klainerman-Nicol\`o \cite{KN} in that initial conditions for $u$ are set on a null hypersurface by means of the area radius parameter and then are extended by geodesics as before. With respect to this coordinate system the candidate Killing field is given by $T=\partial_u$. 
	
	\item \emph{Double-null coordinates.} In order to capture the decaying properties of a function towards both past and future null infinities one is forced to use double-null coordinates. Indeed, Klainerman-Nicol\`o revised the methods of \cite{CK} using a double-null foliation which is better suited for the analysis of the structure equations. \\
	
	The relevance of these coordinates in this paper relies on the fact that the Carleman estimates, Theorem \ref{thm:Carleman}, are stated with respect to double-null coordinates. It is important to stress, however, that the type of metrics allowed by definition \ref{def:AF} do not satisfy a priori the decay conditions, (\ref{metric AS}), required for the Carleman estimates to hold. However, in the context of non-radiating spacetimes, it can be proved that the Cauchy development of initial data close to Kerr-Newman which happens to be non-radiating do admit double-null coordinates of the desired form (\ref{metric AS}). We will prove this in Lemma \ref{lem:doublenull}.
	\end{itemize}

\subsection{Asymptotic conditions}

Now we present the functional spaces relevant for our class of asymptotically flat spacetimes. \\

We say that a function $f \colon \R \times [R, \infty) \times \mathbb{S}^2 \longrightarrow \R$  is in $\mathcal{O}^{\infty}_{k} (r^{-q})$, $q\geq 1$, if it admits an infinite asymptotic expansion in $r^{-n}$, $n=q,q+1,\ldots$, and this expansion is well-behaved with respect to derivatives up to order $k$. Formally we require that there exist functions\footnote{By the symbol $\overset{(n)}{f}$ we mean the function that achieves the desired approximation to $f$ at a given order, as opposed to the $n$th derivative of $f$} $\overset{(n)}{f}(t,\vartheta^i)\in C^k(\R \times \mathbb{S}^2)$ such that for all $t\in \R$, $r\geq R$, $\vartheta^i \in \mathbb{S}^2$ the following holds

	\[ f(t,r,\vartheta^i) \sim \sum_{n=0}^{\infty} \overset{(n-q)}{f} (t,\vartheta^i) r^{q-n} ,\]
	\[ \partial_t^l \partial^{\al}_{\vartheta} f(t,r,\vartheta^i) \sim \sum_{n=0}^{\infty} \partial_t^l \partial^{\al}_{\vartheta} 
	\overset{(n-q)}{f} (t,\vartheta^i) r^{q-n},	\]
	\textrm{ where } $\partial^{\al}_{\vartheta} = \dfrac{\partial^{\al_1}}{\partial_{\vartheta^{1}}} \dfrac{\partial^{\al_2}}{\partial_{\vartheta^{2}}}, \, l + \al_1 + \al_2 \leq k$.
	\[ \partial^l_{r} f(t,r,\vartheta^i) \sim \sum_{n=0}^{\infty} \prod_{i=0}^{l-1} (q-n-i) \overset{(n-q+l)}{f} (t,\vartheta^i) r^{q-n-l}, \, \quad 1\leq l \leq k.\]
	Here $f \sim \sum_{n=0}^{\infty} \overset{(n-q)}{f} r^{q-n}$	means that for all $N\in \N$ there is $C_N>0$ so that
		\[ \big| f - \sum_{n=0}^N \overset{(n-q)}{f} r^{q-n} \big| \leq C_N r^{q-N-1} . \]

\begin{definition} \label{def:AF} 
Throughout this paper we will call a Lorentzian manifold $(\mathcal{M},g)$ \emph{asymptotically flat} if the following conditions are satisfied:
	\begin{enumerate} [i)]	
\item \emph{Regularity at spatial infinity.} We say that $(\mathcal{M}, g)$ is regular at spatial infinity if there are coordinates $(t,r , \vartheta^i)$ such that for large $r$, the metric admits an expansion of the form
	\begin{eqnarray*}
		 g = -dt^2 + dr^2 + r^2 \breve{\gamma}_{ij} d\vartheta^i d\vartheta^j + g^{\infty}	
	 \end{eqnarray*}	
where $\breve{\gamma}$ is the round metric on $\mathbb{S}^2$ and the components $g_{\al\be}^{\infty}$ of $g^{\infty}$ belong to the class $\mathcal{O}^{\infty}_{2} (r^{-1})$. 

\item \emph{Smoothness at null infinity.} $(\mathcal{M},g)$ is called \emph{smooth at future null infinity} if, with respect to the null out-going coordinates $(u,s, \theta^2, \theta^3)$ previously defined, the vanishing of a component of a geometric quantity, $w = w (s, u, \theta^1, \theta^2)$, in the sense that
		\begin{equation}	\label{vanish}
				\lim_{u; r \rightarrow \infty} r^q w = 0, 	
		\end{equation}	
	implies $r^q \omega = \mathcal{O}(r^{-1})$ and the existence of the limit of $w$ to order $q+1$, in the sense that
		\[ \partial_s (r^{q+1} w) = \mathcal{O}(r^{-2}) .	\]
	Moreover, we require a mild angular regularity condition, namely, if $w$ vanishes in the sense of (\ref{vanish}) then so do its angular derivatives:
		\[  	\lim_{u; r \rightarrow \infty} r^q \partial_{\theta^i} w = 0 . 	\]
	Analogously, we require these conditions to hold at past null infinity.
	
\item \emph{Weak peeling.} We also require that the Weyl tensor decays accordingly to
		\begin{align*}
			|\underline{\al} |_{\cancel{g}}	 &= \mathcal{O}(r^{-1}) 	\\
			|\underline{\be} |_{\cancel{g}}	 &= \mathcal{O}(r^{-2}) 	\\
			|\rho|, |\si| 					 &= \mathcal{O}(r^{-3}) 	\\
			|\partial_u \be |_{\cancel{g}}	 &= \mathcal{O}(r^{-4}) 	\\
			|\partial_u \alpha|_{\cancel{g}} 			 &= \mathcal{O}(r^{-5})
		\end{align*}

\item \emph{Compatibility of $I^+$ and $I^-$ at spatial infinity.}	We say that the \emph{poles are finite towards spatial infinity} if all the components of the curvature $\kappa = \al, \be, \rho, \si, \underline{\be}, \underline{\al}$ admit an expansion for large $r$ (near future and past null infinities $I^+$, $I^-$), 
		\[ \ka \sim \sum_{n=0}^{\infty} \overset{(n+1)}{\ka}_{\pm} (u_{\pm}, \xi) r^{-1-n}, \qquad I^{\pm} \simeq \{ (u_{\pm}, \xi): u_{\pm}\in \R, \xi \in \mathbb{S}^2  \}	, \]
		with the property that
		\[	\lim_{u_{\pm} \rightarrow \mp \infty} | \overset{(n)}{\ka_{\pm}} (u_{\pm} ,\xi)| < \infty , \qquad \forall n\in \N.	\]

\end{enumerate}
\end{definition}

These are all technical conditions which are sufficient to allow us to deal with the Einstein equations one order at a time at infinity. Some remarks are in order. 	\\
\\
\textbf{Remarks.} \begin{itemize}
\item The first condition states that the initial regularity assumptions at spatial infinity are propagated through out the evolution. This property may not hold in general since logarithmic divergences are expected to appear when $t\rightarrow \infty$. However, it should be possible to prove its validity for arbitrarily large $t$ by using Friedrich's conformal equations, \cite{Friedrich02}. Then condition $i)$ would hold for time-periodic spacetimes. 

\item The purpose of condition $ii)$ is to rule out \emph{logarithmic singularities}. For example, the function $\frac{\log r}{r}$ decays at infinity but it is not $\order(r^{-1})$. It is known that these kind of singularities are present in Cauchy developments of asymptotically flat initial data \cite{Winicour} and actually are expected to be ubiquitous in dynamical spacetimes. However, as remarked by Christodoulou \cite{Christodoulou02} the coefficients accompanying the logarithmic singularities are time-independent for relevant systems, e.g., an $N$-body system with no incoming radiation. Therefore we can get rid off the logarithmic singularity by taking an extra time-derivative. In the context of non-radiating spacetimes taking an extra time-derivative does not affect the general argument thanks to the \emph{finiteness of poles condition}, condition $iv)$ which prohibits polynomial growth.	\\

Summarising, at first glance it would appear that smoothness at null infinity is too strong an assumption; however, thanks to the CK analysis, we can circumvent the problem by noting that an extra time-derivative makes the geometric quantities smooth at null infinity. The same issue arises when discussing the Peeling property. It does not hold in general for all the components of the Weyl tensor, but when the failure is understood as being time-independent, an extra time-derivative allows to state the weak peeling property as in $iii)$ of the above definition.\\

Moreover, in the context of time-periodic spacetimes, it is easy to prove that the affine parameter $s$ is comparable to $r$ by appealing to a fundamental domain argument and condition $i)$. That is, in a time-periodic spacetime the regularity assumption at null infinity follows from the regularity condition at spatial infinity.\\

At this point it is also worth stressing that the $r$ function appearing in time-space coordinates is also compatible with the $r$ function appearing in double-null coordinates. This is a consequence of Lemma \ref{lem:doublenull}. Therefore, when discussing asymptotic properties, we can freely use any of these $r$ or $s$ as parameters along the outgoing null geodesics.

\item Finally, there are dynamical spacetimes satisfying the above regularity assumptions. Indeed, Cutler and Wald, \cite{Cutler-Wald}, and Chru\'sciel and Delay, \cite{Chrusciel-Delay} provided examples of electrovacuum and vacuum spacetimes, respectively, having a smooth null infinity.  
\end{itemize}	
	
\subsection{Quantities to all orders at infinity} \label{section:quantities}

Following the the work of Bi\v{c}\'ak-Sholtz-Tod \cite{BST} and Alexakis-Schlue \cite{AS} we now proceed to show that the radiation fields $\Xi_{ij}:=\overset{(1)}{\hat{\chi}_{ij}}$ and $\underline{A}(F)_i :=\overset{(1)}{\underline{\al}(F)_i} $ characterise the metric at infinity to all orders (up to stationary data to be described below). To do so we rely on the hierarchy discovered by BMS which is also tied to the splitting of the Weyl tensor into its null components and can be interpreted as signature levels.	\\

Formally, in Proposition \ref{prop:recurrence relations} we state recurrence relations satisfied by the different orders of the physical quantities. This is done by translating the structure equations to infinity, hence obtaining a non-linear algebraic system of equations for the asymptotic coefficients. Then the aforementioned hierarchy helps us to identify levels where the equations become linear for the quantities belonging to that level. Moreover, the structure of that hierarchy leads to identifying the radiation fields as the necessary initial data to run an induction argument. \\

Then we further specialise to the case when $\Xi_{ij}=0$ and $A(F)_i=0$, that is, when no radiation is escaping to infinity (Definition \ref{non-radiating}). Then by running the hierarchy-induction procedure we prove in Corollary \ref{cor:non-radiating} that $\partial_u$ is a Killing symmetry to all orders at infinity.	\\

We remark here that we will be using generically the word \emph{quantities} to refer to the components of either the orthonormalisation matrix, the connection coefficients or the curvature components. \\

Before stating the main results we set some notation. Recall that $\overset{(n)}{f}$ denotes the best $r^{-n}$-approximation of $f$ (see  discussion preceding definition \ref{def:AF}). The symbol $\{\phi_1, \ldots, \phi_n\}$ will denote any expression involving the functions $\phi_1, \ldots, \phi_n$. The symbol $\lfloor n \rfloor $ denotes an expression involving the connection coefficients up to order $n$ and the orthonormalisation matrix up to order $n-1$. That is, 
	$$
	\lfloor n \rfloor  = \{ \overset{(0)}{\omega_{\mu \nu \la}}, \ldots ,  \overset{(n)}{\omega_{\mu \nu \la}},  \overset{(0)}{{h_\mu}^{\nu}},  \ldots  \overset{(n-1)}{{h_\mu}^{\nu}}  \}	,
	$$
with the convention that $\lfloor 0 \rfloor  =0$. The symbol $Q(\phi,\ldots; \varphi, \ldots)$ stands for a quadratic expression containing terms of the form $\phi \varphi$. We remind the reader that asymptotic flatness is to be understood as stated in definition \ref{def:AF}. \\

\begin{prop} \label{prop:recurrence relations}
Let $(M, g, F)$ be an asymptotically flat solution of the Einstein-Maxwell equations. Then the asymptotic quantities satisfy the following recurrence relations along the out-going null hypersurfaces $C_u^+$ for any $n\in \N$,
	\begin{subequations} 	\label{recursion0}
	\begin{align}
	 \overset{(n+1)}{\alpha_{ij}} &= (n-1) \overset{(n)}{\chi_{ij}} - 2 \overset{(n)}{\al(F)_k} \overset{(1)}{\al(F)^k}  \eta_{ij} + \lfloor n-1 \rfloor , 	\\
	 \overset{(n)}{{h_i}^j} &= \overset{(n)}{{\chi_i}^k} \overset{(1)}{{h_k}^j} + \lfloor n-1 \rfloor , 
	\end{align}
	\end{subequations}
	\begin{subequations} \label{recursion1}
	\begin{align}
	 \overset{(n+1)}{\beta_i}, \overset{(n)}{\rho(F)}, \overset{(n)}{\si(F)}, \overset{(n)}{\omega_{jji}}, \overset{(n)}{\zeta_i}, \overset{(n)}{f^i} &= \{ \overset{(n)}{\chi_{ij}}, \overset{(n)}{\al(F)_i}, \lfloor n-1 \rfloor \} ,	\quad  n> 3	
	\end{align}
	\end{subequations}
	\begin{subequations}	\label{recursion2}
	\begin{align}
	 \overset{(n+1)}{\rho}, \overset{(n+1)}{\si}, \overset{(n)}{\underline{\al}(F)}, \overset{(n)}{\underline{\omega}}, \overset{(n)}{\omega_{123}}, \overset{(n)}{\underline{\chi}_{ij}}, \overset{(n)}{\la_i},  \overset{(n-1)}{f^0} &= \{ \overset{(n)}{\chi_{ij}}, \overset{(n)}{\al(F)_i}, \lfloor n-1 \rfloor \} , \quad n> 2 .	
	 \end{align}
	\end{subequations}
Moreover, 
	\begin{subequations}	\label{recursion3}
	\begin{align}
	2\partial_u \overset{(n+1)}{\chi_{ij}} &= -n \overset{(n)}{\underline{\chi}_{ij}} + \{ \overset{(n)}{\chi_{ij}}, \overset{(n)}{\al(F)_i}, \lfloor n-1 \rfloor \} , 		\\
	2\partial_u \overset{(n+1)}{\al(F)_i} &= \{ \overset{(n)}{\al(F)_i} , \overset{(n)}{\rho(F)}, \overset{(n)}{\si(F)}, \lfloor n \rfloor \}, 
	\end{align}
	\end{subequations}
	\begin{subequations} \label{recursion4}
	\begin{align}
	\overset{(n+1)}{\underline{\be}_i} &= \partial_u \overset{(n+1)}{\zeta_i} + \lfloor n \rfloor , 			\\
	\overset{(n+1)}{\underline{\al}_{ij}} &= \partial_u \overset{(n+1)}{\underline{\chi}_{ij}} + \lfloor n \rfloor .
	\end{align}
	\end{subequations}
\end{prop}

This proposition can be interpreted as saying that the asymptotic quantities can be computed recursively starting from the radiation fields. The cases $n=1,2,3$ are special. Roughly speaking $n=1$ corresponds to the choice of gauges at infinity. On the other hand, for $n=2$  we have to specify a mass aspect function as data at the $(\rho, \sigma)$-level; also an EM-charge aspect function at the $\beta_i$-level. Finally, for $n=3$ at the $\be_i$-level an angular momentum aspect vector is required. It is worth remarking that these functions are not freely specifiable but they obey constraint/evolution equations given by the Einstein and Maxwell equations of signature $-1$ and $-2$. We deal with these cases during the proof of Proposition \ref{prop:recurrence relations}. Here we state the main relations found there in:
 \begin{lem} \label{lem:evolution equations}
	Under the same hypothesis of Proposition \ref{prop:recurrence relations} the following relations hold,
	\begin{subequations}	\label{mass and angular momentum}
	\begin{align}
	\partial_u \tr \overset{(2)}{\underline{\chi}} &= Q(\overset{(1)}{\underline{\chi}}; \overset{(1)}{\underline{\chi}}) + Q(\overset{(1)}{\underline{\al}(F)}; \overset{(1)}{\underline{\al}(F)}),	\\
	 \overset{(3)}{\rho}, \overset{(3)}{\si}, \overset{(2)}{\underline{\al}(F)}, \overset{(2)}{\underline{\omega}}, \overset{(2)}{\omega_{123}}, \overset{(2)}{\hat{\underline{\chi}}_{ij}}, \overset{(2)}{\la_i} , \overset{(1)}{f^0} &= \{ \overset{(2)}{\chi_{ij}}, \overset{(2)}{\al(F)_i}, \tr \overset{(2)}{\underline{\chi}}, \lfloor 1 \rfloor \} ,	\\
 		 	\overset{(3)}{\omega}_{jji} &= \overset{(1)}{h^{-1}}  \cdot ( \breve{e}_k \overset{(3)}{{h_i}^j} + \lfloor 2 \rfloor ), \\
	 	\partial_u\overset{(3)}{\zeta_i} &= - \partial_u\overset{(3)}{\omega_{jji}} + Q(\overset{(1)}{{ \underline{\al}(F)}}; \overset{(2)}{\rho(F)}, \overset{(2)}{\si(F)}) + \lfloor 2 \rfloor  , \\
	 	2 \partial_u\overset{(2)}{\rho(F)} &=  \breve{\cancel{\divergence}} \, \underline{A}(F) , \\
	 	2 \partial_u\overset{(2)}{\sigma(F)} &= \breve{\cancel{\curl}} \, \underline{A}(F) .
	\end{align}
	\end{subequations}
 \end{lem}
Here $\breve{e}_k := \overset{(1)}{{h_k}^j} \partial_j$ is the standard orthonormal basis on the round sphere and $\breve{\cancel{\divergence}}$ and $\breve{\cancel{\curl}}$ are the corresponding operators. We remark that equations (\ref{mass and angular momentum}a), (\ref{mass and angular momentum}d), (\ref{mass and angular momentum}e) and (\ref{mass and angular momentum}f) can be interpreted as evolution formulas for the mass, angular momentum and EM-charges. From the point of view of a characteristic initial value formulation, these are constraint equations at null infinity.	\\
 
In particular, combining these relations with the non-radiating condition we get stationarity to all orders at infinity: \\

\begin{cor}  \label{cor:non-radiating} 
Let $(M, g, F)$ be an asymptotically flat electrovacuum spacetime. Assume it is non-radiating, that is, $\Xi_{ij}=0=\underline{A}(F)_{i}$. Then all the asymptotic quantities are $u$-independent.	\\
\end{cor}

Along the same lines we can prove a related result, namely, that all the asymptotic quantities can be found recursively in terms of the radiation fields and the ``pole moments'' of the initial Cauchy data. In other words, the radiation fields determine completely the dynamics of the spacetime at infinity. \\

\begin{cor}
Let $(M, g, F)$ be an asymptotically flat electrovacuum spacetime. Then all the asymptotic quantities depend solely on the radiation fields $\Xi_{ij}$, $\underline{A}(F)_{i}$, the initial mass and angular momentum and the pole moments of $\chi_{ij}$ and $\al(F)_i$ (The latter are the limits as $u\rightarrow -\infty$ of the asymptotic components $\overset{(n)}{\chi_{ij}}$ and $\overset{(n)}{\al(F)_i}$, $n \in\N$). 
\end{cor}

Before proving these results we present the structure equations below. We have grouped them in levels representing the hierarchy found first by BMS, \cite{BMS}, \cite{Sachs}. Note that (\ref{hierarchy0}b), (\ref{hierarchy1}e), (\ref{hierarchy1}f), (\ref{hierarchy2}i) and (\ref{hierarchy2}j) are frame equations while (\ref{hierarchy1}g), (\ref{hierarchy2}k) are 1st Bianchi identities. All of these are satisfied on any Lorentzian manifold. In contrast, the Einstein equations appear implicitly when we regard $S_{\mu\nu}$ as a quadratic expression in $F_{\mu \nu}$. Finally, (\ref{hierarchy1}c), (\ref{hierarchy1}d), (\ref{hierarchy2}g), (\ref{hierarchy2}h), (\ref{hierarchy3}c), (\ref{hierarchy3}d) are the  Maxwell equations.

\begin{enumerate}	[a)]
\item The $\alpha_{ij}$ or $\Psi_0$ or signature $2$ level. This includes the 2nd fundamental form $\chi_{ij}$ and ${h_i}^j$:	\begin{subequations}	\label{hierarchy0}
	\begin{align}	
		\alpha_{ij}	+ \frac12 \eta_{ij} S_{00} &= -e_0(\chi_{ij}) - \chi_{ik} {\chi^k}_j , \\
		e_0({h_i}^j)& = -{\chi_i}^k {h_k}^j.
	\end{align}
	\end{subequations}
	
\item The $\beta_i$ or $\Psi_1$ or signature $1$ level. This includes the torsion $\zeta_i$ and the coefficients of the induced connection on $S_{s,u}$, that is, $\omega_{iij}$, and $f^i$:
	\begin{subequations}	\label{hierarchy1}
	\begin{align}
		2\beta_i + S_{0i} &= -2 e_0(\zeta_i) - 4 \zeta^k \chi_{ki}	, \\
		 \beta_i - \frac12 S_{0i} &= e_0(\omega_{jji}) -  \zeta_j\chi_{ii} + \chi_{jj}\zeta_i + {\chi_j}^k\omega_{kji} , \quad i\neq j , 	\\
		e_0(\rho(F)) &= - \cancel{\divergence} \, \al(F) - \tr \chi \rho(F) - \zeta_i\al(F)^i , 	\\
		e_0(\si(F)) &= - \cancel{\curl} \, \al(F) - \tr \chi \si(F) + \varepsilon^{ik}\zeta_i\al(F)_k , 	\\
		e_i({h_j}^k) - e_j({h_i}^k) &= ({\omega_i}^n{}_j - {\omega_j}^n{}_i) {h_{n}}^k, 		\\
		e_0(f^i) &= -\zeta^k {h_k}^i , \\
		e_1(\omega_{jji}) - e_j(\omega_{1ji}) &= -e_j(\underline{\chi}_{ji}) + e_i(\underline{\chi}_{jj}) \nonumber 
										\\	& \qquad - (\omega\wedge\omega)_{ji1j} +  (\omega \wedge \omega)_{1jji} , \quad i\neq j .
	\end{align}		
	\end{subequations}
with no summation on repeated $j$'s. 

\item The $(\rho, \si)$ or $\Psi_2$ or signature $0$ level. This includes $\underline{\omega}$, $\omega_{123}$, $\underline{\chi}_{ij}$, $\la_i$ and $f^0$:
	\begin{subequations} \label{hierarchy2}
	\begin{align}
		-\rho + \frac12(S_{22} + S_{33}) &= e_2(\omega_{323}) - e_3(\omega_{223}) + \frac12 \chi_{22} \underline{\chi}_{33} + \frac12 \chi_{33} \underline{\chi}_{22} \nonumber 
										\\	& \qquad  -  \chi_{23} \underline{\chi}_{23} - (\omega_{232})^2 - (\omega_{323})^2 , \\
		- \si &= -e_2(\zeta_3) + e_3(\zeta_2) + \omega_{323}\zeta_3 - \omega_{232}\zeta_2  \nonumber 
										\\	& \qquad 	 - {\chi_2}^k \underline{\chi}_{3k} + 	{\chi_3}^k \underline{\chi}_{2k} 		, \\
		 \rho +\frac12 S_{01} &= e_0(\underline{\omega}) - 3\zeta^k\zeta_k, 	\\
		- 2 \sigma &= e_0(\omega_{123}) + 4 \zeta^k \omega_{k23}	, 	\\
		-4 \rho - S_{jj} + \frac12 S_{01} &= -e_0(\underline{\chi}_{jj}) - 2 e_j(\zeta_j) + 2 \zeta^k \omega_{jkj} - {\chi_j}^k\underline{\chi}_{kj} - 2\zeta_j \zeta_j, 	\\
		- \sigma - S_{23} &= -e_0(\underline{\chi}_{23}) - 2 e_2(\zeta_3) + 2 \zeta^k\omega_{2k3} - 2 \zeta_2\zeta_3 - {\chi_2}^k\underline{\chi}_{k3} , 	\\
		e_0(\underline{\al}(F)_i) &= -e_i(\rho(F)) + {\varepsilon_i}^je_j(\si(F)) - (F\wedge \omega)_{i}	,	\\
		e_1(\al(F)_i) &= e_i(\rho(F)) + {\varepsilon_i}^k e_k(\si(F)) - ({}^* F\wedge \omega)_{i}	,	\\
		e_0(f^0) &= \underline{\omega}, \\
		e_i(f^0) &= \la_i	, \\
		e_1(\chi_{ij})	-e_i(\zeta_j)	 &=  e_0(\underline{\chi}_{ij}) - e_j(\zeta_i) - (\omega \wedge \omega)_{1i0j} +  (\omega \wedge \omega)_{0j1i} .
	\end{align}		
	\end{subequations}
	With no summation on repeated $j$'s. 
\item The $\underline{\beta}_i$ or $\Psi_3$ or signature $-1$ level:
	\begin{subequations} \label{hierarchy3}
	\begin{align} 
	2 \underline{\beta}_i - S_{1i} &= e_0(\la_i) +2 e_1(\zeta_i) - (\omega\wedge \omega)_{01i1} \\
	- \underline{\beta}_i - \frac12 S_{1i} &= e_1(\omega_{jji}) - e_j(\omega_{1ji}) - (\omega \wedge \omega)_{1jji} . \\
		e_1(\rho(F)) & =  \cancel{\divergence} \, \underline{\al}(F) + \tr \underline{\chi} \rho(F) + \zeta_i\al(F)^i , 	\\
		e_1(\si(F)) & = -\cancel{\curl} \, \underline{\al}(F) - \tr \underline{\chi} \si(F) + \varepsilon^{ik}\zeta_i\al(F)_k , 	
	\end{align}
	\end{subequations}
With no summation on repeated $j$'s. 

\item Finally, the $\underline{\al}_{ij}$ or $\Psi_4$ or signature $-2$ level:
\begin{align} \label{hierarchy4}
	\underline{\al}_{ij} + \frac12 \eta_{ij}S_{11} = -e_i(\la_i) - e_1(\underline{\chi}_{ij}) - (\omega\wedge \omega)_{i1j1}. 
\end{align}
\end{enumerate}

Now we proceed to prove the results stated above. It is convenient to think of the structure-Einstein-Maxwell equations as a 1st order system schematically of the form:
		\begin{align*}
		  \nabla h + \omega \wedge h &= 0 , \\
		 \nabla \omega + \omega \wedge \omega &= C + \eta \wedge S , \\
		 S &= F^2, \\
		 \nabla F + \omega \wedge F &= 0 .
		\end{align*}
The procedure will be analogous to the one described in the Introduction for a wave equation. \\

\textbf{Proof of Proposition \ref{prop:recurrence relations}.} Firstly, we state the 1st order values of the relevant coefficients that we will be using to obtain the recurrence relations and also in the induction argument. These are a consequence of the CK analysis and the gauge conditions. 
		\begin{align*}
		 \overset{(0)}{f^0}	&= 1  , \qquad f^1 \equiv 2	, \qquad \overset{(1)}{{h_i}^j}	= \mathrm{diag}(1 , \frac{1}{\sin \theta}),	\\
			\overset{(1)}{\underline{\chi}_{ij}} & = \Xi_{ij} + \eta_{ij},	\qquad 	\overset{(1)}{\chi_{ij}} = \eta_{ij}.
		\end{align*}
 	Recall the Einstein equations:
	$$
	S_{\mu \nu } = 2 F_{\mu \si} {F^{\si}}_{\nu} - \frac12 \eta_{\mu \nu} F_{\al \be} F^{\al \be} . 
	$$
They are quadratic in $F_{\mu \nu}$, in particular then $\overset{(n+1)}{S_{\mu \nu}} = \{ \overset{(1)}{F_{\mu \nu}}, \ldots, \overset{(n)}{F_{\mu \nu}} \}$. A more detailed null decomposition is given by
	\begin{align*}
		S_{00} &= 2 \al(F)_i \al(F)^i  = \order (r^{-5}), \\
		S_{01} &=  \rho(F)^2 + \si(F)^2  = \order (r^{-4}), \\
		S_{ij} &= Q (\rho(F), \si(F) ; \rho(F), \si(F)) +  Q(\al(F); \underline{\al}(F)) = \order(r^{-\frac72})	, \\
		S_{0i} &= Q (\al(F)_i ; \rho(F), \si(F)) = \order(r^{-\frac92}), \\
		S_{1i} &= Q (\underline{\al}(F)_i ; \rho(F), \si(F)) = \order(r^{-3}) \\
		S_{11} &= 2 \underline{\al}(F)_i \underline{\al}(F)^i = \order(r^{-2})  , 
	\end{align*}	 
where we have included the peeling behaviour of the Faraday tensor, (\ref{null Faraday2})-(\ref{null Faraday3}). Now we proceed to derive the recurrence relations. 

\begin{enumerate}	[a)]
\item It can be seen that the $(n+1)$-order of the $\al$-level equations together with 
$$ \overset{(n+1)}{S_{00}} =  \overset{(n)}{\al(F)_i} \overset{(1)}{\al(F)^i} + \{ \overset{(1)} \al(F)_i, \ldots, \overset{(n-1)} \al(F)_i \} $$
imply the recurrence relations (\ref{recursion0}) for $\overset{(n+1)}{\al_{ij}}$ and $\overset{(n)}{{h_i}^j}$. 

\item Now we look at the $(n+1)$-order of the $\be$-level equations. From this level onwards the equations become more intricate. The important thing to remember is that we only need to keep track of the coefficients accompanying the variables that we want to find at a given order and level. \\

We start with the $(n+1)$-order of the frame equation:
	$$e_i({h_j}^k) - e_j({h_i}^k) = ({\omega_i}^n{}_j - {\omega_j}^n{}_i) {h_{n}}^k . $$
Note that the left-hand-side does not contain terms with $\overset{(n+1)}{{h_j}^k}$ since $e_i = {h_i}^j\partial_j$ and ${h_i}^j$ decays. Writing down the equations explicitly for $k=2,3$ we get the system
		\begin{eqnarray*}
	 \left( \begin{array}{cc}
	\overset{(1)}{{h_2}^2} & - \overset{(1)}{{h_2}^3}\\
	-\overset{(1)}{{h_2}^2} &  \overset{(1)}{{h_2}^3}
 \end{array}\right) 
 	 \left( \begin{array}{c}
	\overset{(n)}{\omega_{223}} \\
	\overset{(n)}{\omega_{332}}
 \end{array}\right) = \breve{e}_k \overset{(n)}{{h_i}^j} +  \lfloor n-1 \rfloor . 
\end{eqnarray*}
Here we recall that $\breve{e}_k := \overset{(1)}{{h_k}^j} \partial_j$ is the standard orthonormal basis on the unit round sphere. This system tells us that $\overset{(n)}{\omega_{jji}} = \{ \breve{e}_k \overset{(n)}{{h_i}^j},  \lfloor n-1 \rfloor \} = \{ \breve{e}_k \overset{(n)}{\chi_{ij}},  \lfloor n-1 \rfloor \}$, where we have used the recurrence relation for $\overset{(n)}{{h_i}^j}$. In particular for $n=3$ we get equation (\ref{mass and angular momentum}c).\\

We are left with the variables $\be_i$, $\rho(F)$, $\si(F)$ and $\zeta_i$. The $(n+1)$-order of equations (\ref{hierarchy1}a)-(\ref{hierarchy1}d) give the following linear system valid for $n\geq 3$, 
		\begin{eqnarray*}
	 \left( \begin{array}{cccc}
	2 & 0 & 0 & -(2n-4) \\
	1 & 0 & 0 & -1 \\
	0 & -(n-2) & 0 & 0 \\
	0 & 0 & -(n-2) & 0 
 \end{array}\right) 
 	 \left( \begin{array}{c}
	\overset{(n+1)}{\beta_i}	 \\
	\overset{(n)}{\rho(F)}	 \\
	\overset{(n)}{\si(F)}	 \\
	\overset{(n)}{\zeta_i}	 
 \end{array}\right) =  \{ \overset{(n)}{\chi_{ij}},\overset{(n)}{\al(F)_i},  \lfloor n-1 \rfloor \} , 
\end{eqnarray*}
where the zeros on the first two rows correspond to the fact that $\overset{(1)}{\al(F)} = 0 = \overset{(2)}{\al(F)} $ and the quadratic character of $S_{0i}$ stated before. Similarly the zeros on the last column come also from $\overset{(1)}{\al(F)} = 0 = \overset{(2)}{\al(F)}$. The above system can be solved (note that the degenerate cases correspond to $n =2,3$) to obtain the desired recurrence relations. \\

It is worth noticing that $\overset{(1)}{\zeta_i} = 0$, while $\overset{(1)}{\omega_{jji}}$ corresponds to the connection coefficients of the standard round sphere.\\
\\
\textbf{Remark.} The case $n=2,3$ are special. For $n=2$ the Maxwell equations degenerate and we have to specify $\overset{(2)}{\rho(F)}$ and $\overset{(2)}{\sigma(F)}$ as initial data, these can be regarded as electromagnetic charges. For $n=3$, the degeneracy is telling us that we have to prescribe $\overset{(3)}{\zeta_i}$ or $\overset{(4)}{\be_i}$ as initial data. However, these data are not freely specifiable, in the sense that they satisfy evolutions equations (\ref{mass and angular momentum}). We will find the same situation on the $(\rho,\si)$-level where the corresponding initial data can be regarded as a mass aspect function.

\item This is the 0-signature case; a signature count gives us that $\la_i$ does not appear on the $(\omega \wedge \omega)$-term. Thus, the $(n+1)$-order of equations (\ref{hierarchy2}a)-(\ref{hierarchy2}g) give us 9 equations for 9 variables, namely, $\overset{(n+1)}{\rho}$, $\overset{(n+1)}{\si}$, $\overset{(n)}{\al(F)_i}$, $\overset{(n)}{\underline{\omega}}$, $\overset{(n)}{\omega_{123}}$ and $\overset{(n)}{\underline{\chi}_{ij}}$. The linear system can be solved provided it is non-degenerate, which can be checked by direct computation for $n>2$. This gives the desired recurrence relation.	Here we state the linear system obtained by considering the $(n+1)$-order of equations (\ref{hierarchy2}a)-(\ref{hierarchy2}g):
		\begin{eqnarray*}
	 \left( \begin{array}{cccccccc}
	1 & 0 & \frac12 & \frac12 & 0 & 0 & 0 & * \\
	0 & 1 & 0 & 0 & 0	& 0	& 0 & 0 \\
	1 & 0 & 0 & 0 & 0	& n	& 0 & * \\
	0 & 1 & 0 & 0 & 0	& 0	& -n & 0\\
	1 & 0 & n-1 & 0 & 0	& 0	& 0 & * \\
	1 & 0 & 0 & n-1 & 0	& 0	& 0 & *\\
	0 & 1 & 0 & 0 & 0	& 0	& n-1 & *\\
	n-1 & 0 & 0 & 0 & 0	& 0	& 0 & 0\\
 \end{array}\right) 
 	 \left( \begin{array}{c}
	\overset{(n+1)}{\rho}	 \\
	\overset{(n+1)}{\si}	 \\
	\overset{(n)}{\underline{\chi}_{22}}	 \\
	\overset{(n)}{\underline{\chi}_{33}}	 \\
	\overset{(n)}{\underline{\chi}_{23}}	 \\
	\overset{(n)}{\underline{\omega}}	 \\
	\overset{(n)}{\underline{\omega}_{123}}	 \\
	\overset{(n)}{\underline{\alpha}(F)_i}	 \\
 \end{array}\right) = \textrm{Known data} , 
\end{eqnarray*}
where `Known data' can be described more precisely as a term of the form 
$$ \{ \overset{(n)}{\chi_{ij}}, \overset{(n)}{\zeta_i},\overset{(n)}{\omega_{jji}}, \overset{(n)}{\al(F)_i}, \overset{(n)}{\rho(F)},\overset{(n)}{\si(F)},  \lfloor n-1 \rfloor \} .$$

Once again the case $n=2$ is special. The 1st, 5th and 6th rows become linearly dependent. Due to this degeneracy we have to specify $\tr \overset{(2)}{\underline{\chi}}$ along null infinity, this corresponds to the Bondi mass (up to gauge terms). Finally we remark that the Einstein equation at the $\underline{\al}$-level provides us with an evolution equation, (\ref{mass and angular momentum}a), for $\tr \overset{(2)}{\chi}$, which corresponds to the Bondi mass formula (\ref{Bondi mass formula}).

\item Finally, equations (\ref{recursion3}) are obtained by taking the $(n+1)$-order of equations (\ref{hierarchy2}k) and (\ref{hierarchy2}h), respectively.

\item Now we focus on Lemma \ref{lem:evolution equations}. The evolution equations for the EM-charges, (\ref{mass and angular momentum}e) and (\ref{mass and angular momentum}f), follow from Maxwell's equations (\ref{hierarchy3}c) and (\ref{hierarchy3}d), respectively.

\item As explained before we can obtain an evolution equation for $\overset{(3)}{\zeta_i}$ by considering the system $(\ref{hierarchy3}a)$ and $(\ref{hierarchy3}b)$. The equation obtained after getting rid of the Weyl terms is of the form:
	$$
	\partial_u\overset{(3)}{\zeta}_i = - \partial_u \overset{(3)}{\omega_{jji}} +  Q(\overset{(1)}{\underline{\chi}}; \overset{(2)}{\zeta}) + Q(\overset{(1)}{\underline{\al}(F)};\overset{(2)}{\rho(F)}, \overset{(2)}{\si(F)}) + \lfloor 2 \rfloor .
	$$
This proves the identity (\ref{mass and angular momentum}d). 

\item Again a signature analysis as above helps us to find the structure of the $\underline{\al}$-equation. We have that to 2nd order the trace of (\ref{hierarchy4}) reads
	$$
	\partial_u \tr \overset{(2)}{\underline{\chi}} = Q(\overset{(1)}{\underline{\chi}}; \overset{(1)}{\underline{\chi}}) + Q(\overset{(1)}{\underline{\al}(F)}; \overset{(1)}{\underline{\al}(F)}) , 
	$$
where we have used $\overset{(1)}{\zeta_i}=0$. The remaining system for $\overset{(3)}{\rho}$, $\overset{(3)}{\sigma}$, etc, is non-degenerate and can be solved. \\

This finishes the proof of Proposition \ref{prop:recurrence relations} and Lemma \ref{lem:evolution equations}. $_{\blacksquare}$\\
\end{enumerate}

\textbf{Proof of Corollary \ref{cor:non-radiating}.} All the hard work was done in the proof of Proposition \ref{prop:recurrence relations}. Now we have to check the lower order cases. \\

Firstly, we show that the 1st order asymptotic quantities are all $u-$independent. Recall that $\overset{(1)}{{h_i}^j}=\mathrm{diag(1, 1/\sin\theta)}$ and $\overset{(1)}{\chi_{ij}}=\eta_{ij}$. Moreover, $\overset{(1)}{\underline{\al}(F)_i}=0$ and $\overset{(1)}{\underline{\chi}_{ij}}=\eta_{ij}$ by the non-radiating condition. We also know that $\overset{(n)}{\al_{ij}}= \overset{(n)}{\be_i}=0$ for $n=0,1,2,3$; $\overset{(n)}{\al(F)_i} = \overset{(n)}{\rho}= \overset{(n)}{\si}=0$ for $n=0,1,2$; and $\overset{(n)}{\rho(F)}= \overset{(n)}{\si(F)}=0$ for $n=0,1$.\\

Now we focus on $\overset{(1)}{\zeta_i}$ and look at equations (\ref{hierarchy3}), we can get rid of the Weyl term by subtracting them. To 1st order we find that 
	$$ \partial_u \overset{(1)}{\zeta_i} = \{ \overset{(1)}{S_{1i}} \} , $$
which vanishes since $S_{1i} = \order(r^{-3})$ due to Einstein-Maxwell equations. Hence $\overset{(1)}{\zeta_i}$ is also $u$-independent. The frame equation $e_0(f^i) = -\zeta^k {h_k}^i$ to 2nd order gives us that $\overset{(1)}{f^i} = \overset{(1)}{\zeta^k} \overset{(1)}{{h_i}^i}$ and thus it is also $u$-independent. This finishes the $\be$-level.\\

Moving on to the $(\rho,\si)$-level: Recall that $\overset{(1)}{\underline{\chi}_{ij}} = \Xi_{ij} + \eta_{ij}$. By assumption $\Xi_{ij} =0$, so $\overset{(1)}{\underline{\chi}_{ij}}$ is $u$-independent. Moreover, $\overset{(2)}{\rho} =\overset{(2)}{\si} = \overset{(2)}{S_{01}} =0$, then the structure equations (\ref{hierarchy2}c) and (\ref{hierarchy2}d) to 2nd order give us that $\overset{(1)}{\underline{\omega}}$ and $\overset{(1)}{\omega_{123}}$ depends on previous data, so they must be $u$-independent. Finally, to 1st order, the frame equations $ e_i(f^0) = \la_i$ tell us that $\overset{(1)}{\la_i}=0$, since  $\overset{(0)}{f^0}=1$. Therefore all the 1st order quantities at this level are $u$-independent.\\

We will see that we can repeat the argument now for asymptotic quantities of order 2. Firstly, the $u$-derivative of the recurrence relation (\ref{recursion3}) for $n=1$ tell us that 
	\begin{align*}
	\partial_u^2\overset{(2)}{\chi_{ij}} &= \partial_u^2\overset{(2)}{\al(F)_i}, \\
	\partial_u^2\overset{(2)}{\al(F)_i} &= 0 , 
	\end{align*}
where we have used that the 1st order quantities are $u$-independent. Now we make use of the hypothesis that $\overset{(3)}{\al_{ij}}$ and $\overset{(2)}{\al(F)_i}$ must remain finite\footnote{At this point is not really necessary to use this hypothesis since we already know that $\overset{(2)}{\al(F)_i}=0$. We have phrased it this way for the sake of generality in the induction argument.} as $u \rightarrow -\infty$ to rule out linear growth of $\overset{(2)}{\chi_{ij}}$ and $\overset{(2)}{\al(F)_i}$. Hence, $\overset{(2)}{\chi_{ij}}$ and $ \overset{(2)}{\al(F)_i}$ are $u$-independent. \\

Now, at the $\be$-level we find that the equation (\ref{hierarchy1}a) degenerates at this order; the $\overset{(2)}{\zeta_i}$-terms cancel. Also $\overset{(3)}{S_{0i}}=0$, this together with equation (\ref{hierarchy1}b) gives a relation of the form:
		$$\overset{(2)}{\zeta_i} = \overset{(2)}{\omega_{jji}}. $$
On the other hand we can use the frame equation (\ref{hierarchy1}e) as before to find $\overset{(2)}{\omega_{jji}}$ in terms of $\overset{(2)}{{h_i}^j}$ and 1st order terms. This proves that $\overset{(2)}{\zeta_i}$ and $\overset{(2)}{\omega_{jji}} $ are $u$-independent.	\\

As forecast, the Maxwell equations (\ref{hierarchy1}c) and (\ref{hierarchy1}d) also degenerate. This suggests that we need to provide $\overset{(2)}{\rho(F)}$ and $\overset{(2)}{\si(F)}$ as initial data. However, equations (\ref{mass and angular momentum}e) and (\ref{mass and angular momentum}f) impose evolution equations for these. Therefore they are $u$-independent in the non-radiating case. This finishes the $\be$-level at this order.\\

Next, as seen in part c) of the proof of Proposition \ref{prop:recurrence relations}, the equations at the $(\rho, \si)$-level also degenerate. Quick inspection and the fact that $\overset{(3)}{S_{01}}=\overset{(3)}{S_{ij}}=0$ lead to identifying $\overset{(2)}{\tr \underline{\chi}}$ as initial data. Then all the remaining quantities can be computed in terms of this. \\

Moreover, the evolution equation for $\overset{(2)}{\tr \underline{\chi}}$ implies that it is $u$-independent in the non-radiating case. Therefore all the quantities at this level are also $u$-independent.	\\

Thus we have established $u$-independence of all the quantities to 1st and 2nd order. In general, we can prove that the 1st and 2nd order quantities depend solely on the radiation fields and three aspect functions: mass and EM-charges.	\\

It is clear now how to repeat the procedure inductively for $n>2$: Assume that $\overset{(n)}{\chi_{ij}}$, $\overset{(n)}{\al(F)_i}$ and $\lfloor n-1 \rfloor$ are $u$-independent then run the hierarchy-recursion argument to conclude that $\lfloor n \rfloor$ is $u$-independent. \\

At this point we encounter one difficulty when analysing the $\be$-level to order 3. It degenerates, but we already know that we have to prescribe an angular momentum aspect vector $\overset{(3)}{\zeta_i}$ which comes with an evolution equation (\ref{mass and angular momentum}d). We now that we can compute $\overset{(3)}{\omega_{jji}}$ from previous data, equation (\ref{mass and angular momentum}c), hence it is $u$-independent in the non-radiating case. Therefore the evolution equation for $\overset{(3)}{\zeta_i}$ implies that it is also $u$-independent and the induction works through. \\

Finally, the evolution equations for $\overset{(n+1)}{\chi_{ij}}$ and $\overset{(n+1)}{\al(F)}$, (\ref{recursion3}a) and (\ref{recursion3}b),  ensure that they will also be $u$-independent (provided they have finite limits as $u \rightarrow -\infty$) and the inductive step is proved. $_{\blacksquare}$

\section{Unique continuation from infinity }
\setcounter{footnote}{1}

In the previous section we showed that all the asymptotic quantities depend only on $\chi_{ij}$ along the out-going null hypersurface $C_0^+$ and the radiating fields $\Xi_{ij}$ and $\underline{A}(F)_i$. In particular, if $\Xi_{ij}$ and $\underline{A}(F)_i$ vanish then all the quantities are $u$-independent; that is, $T=\partial_u$ is a Killing field to all orders at infinity. The goal now is to extend this symmetry into the spacetime. We do this in Proposition \ref{prop:unique continuation2}, which in conjuction with Corollary \ref{cor:non-radiating}, allows us to prove our main result, Theorem \ref{thm1}.\\

We start here by presenting the necessary techniques to extend the time-like symmetry into a neighbourhood of infinity. The motivation comes from successful applications of the so-called Carleman estimates to prove uniqueness of solutions in the context of hyperbolic equations with pseudo-convex boundary conditions. For example, Ionescu-Klainerman used this approach to prove local unique extension of Killing vector fields across a pseudo-convex hypersurface in \cite{IK}. Also, in \cite{AIK}, Alexakis-Ionescu-Klainerman showed uniqueness of smooth stationary black holes for small perturbations of Kerr by proving the unique continuation property for the Simon-Mars tensor. Here we present the main results when the boundary condition imposed is that of asymptotic flatness. Alexakis-Schlue-Shao showed in \cite{ASS} that linear waves satisfy the unique continuation from infinity property provided they decay faster than any polynomial. The main technical tool is the new Carleman estimates they derived in the context of asymptotically flat spacetimes, we include them here as Theorem \ref{thm:Carleman}.\\

In the second part of this section we revise, and extend to the non-vacuum case, the tensorial equations satisfied by the deformation tensors $\Lie_Tg$, $\Lie_TC$ and $\Lie_T F$ obtained by \cite{IK}. These take the form of transport equations for $\Lie_T g$ and its first derivatives along the outgoing null direction and wave equations for $\Lie_TC$ and $\Lie_T F$. The latter are a consequence of the usual wave equations satisfied by $C$ and $F$, which in turn are implied by the Bianchi and Maxwell equations.\\

Then, in the third section we use Cartesian coordinates to cast the equations in a suitable form (with ``fast decaying coefficients'') such that the Carleman estimates from Theorem \ref{thm:Carleman} can be applied. Finally, a standard argument is used to bound the weighted $L^2$-norms of the deformation tensors and hence conclude its vanishing in a neighbourhood of infinity. Special care has to be taken regarding the wave equation for $\Lie_T F$ whose coefficients do not decay fast enough. To deal with this problem, we ``borrow'' some decay from the coupling coefficient appearing in the wave equation for $\Lie_T C$.

\subsection{Carleman estimates} \label{section:Carleman estimates}

In \cite{AS}, Alexakis and Schlue proved Theorem \ref{thm1} for a vacuum spacetime. In order to generalise the proof to include a Maxwell field we need to adapt their argument at the level of Carleman estimates. These are inequalities that can be thought of as a priori estimates for functions decaying faster than any polynomial at infinity. In conjunction with a wave equation satisfied by the function, this method can be used to prove the vanishing of the function in a neighbourhood of infinity.	\\

We present here the Carleman estimates obtained by Alexakis, Schlue and Shao in \cite{ASS} for spacetimes of positive mass. To do so we introduce some notation first.	\\

The class of spacetimes considered in \cite{ASS} are incomplete 4-manifolds of the form $(-\infty,0) \times (0, \infty) \times \mathbb{S}^2$ with coordinates $(u,v, y^i)$ equipped with a metric given by:
	\begin{align} \label{metric AS}
	 g & = 	g_{uu}du^2 - 4 K du dv + g_{vv} dv^2 + \sum_{i.j=2}^3 \gamma_{ij} dy^i dy^j 	\\	\nonumber
	 	& \qquad	\sum_{i=2}^3 (g_{iu} dy^i du + g_{iv} dy^i dv) ,
	\end{align}
where $m$ and $r$ are smooth positive functions. In addition Alexakis-Schlue-Shao require the following bounds to hold:
\begin{itemize}
	\item The metric components satisfy	
	\begin{align*}
	K & = 1- \frac{2m}{r} ,\qquad	& 	g_{uu}, g_{vv} &= \order_1(r^{-3}),	\\
	g_{iu}, g_{iv} &= \order_1(r^{-1})	, \qquad &	\gamma_{ij} &= \breve{\gamma}_{ij} + \order_1(r^{-1}).
	\end{align*}
	\item The function $m$ is uniformly bounded away from $0$, $m\geq m_{min} >0$, and for some $\eta>0$, the differential of $m$ satisfies
	\begin{align*}
	|\partial_i m| = \order(r(-uv)^{-\eta}), \qquad	|\partial_um|, |\partial_v m| = \order(r^{-2}).
	\end{align*}
	\item  The function $r$ is bounded on a level set of $v-u$, that is, there exist constants $C_1$, $C_2$ such that
	\begin{align*}
	v(p)-u(p) = C_1 \quad \textrm{implies} \quad	1 \ll |r(p)| < C_2 .
	\end{align*}
	The differential of $r$ satisfies the following estimate
	\begin{align*}
	\left(	1+ \frac{2m}{r}\right) dr = (1+\order(r^{-2})) dv - (1+\order(r^{-2}))du + \sum_{i=2}^3 \order(r^{-1})dy^i.
	\end{align*}
	Note that these two conditions imply that $r$ and $v-u$ are comparable, that is, there exists a constant $C$ such that $r \leq C |v-u|$.
	\item Finally they also require
	\begin{align*}
	|  \square_g \left( \frac{m}{r}\right)| = \order((-uv)^{-1-\eta}).
	\end{align*}
\end{itemize}

\textbf{Remark.} It is not clear that an asymptotically flat spacetime in the sense of definition \ref{def:AF} satisfies the above condition when cast in double-null coordinates. While this may not be true in general, it holds for non-radiating spacetimes. We prove this in Lemma \ref{lem:doublenull}.	\\

\begin{figure} \label{fig:pseudo-convex}
		\begin{minipage}{8cm}
			\begin{center}
			\begin{tikzpicture}[scale=1]		
	\draw (-3,0) node {};
	\fill (0,4) circle (2pt) node[above] {$i^+$};
	\fill (0,-4) circle (2pt) node[below] {$i^-$};
	\fill (4,0) circle (2pt) node[right] {$i^0$};

	\draw [line width = 1.5pt, red](1.7,2.3) ..controls (2.7,0).. (1.7,-2.3);
	\draw [red]		(1.7,2.3)  (0,0.6);
	\draw [red]		(1.7,-2.3)  (0,-0.6);

	\draw [line width = 1.5pt, dashed] (0,-4) -- (0,4) node[pos=.5,sloped,above] {$r=0$};
	\draw [line width = 1.5pt] (0,4) -- (4,0) node[pos=.7, above right]{$I^+$};
	\draw [line width = 1.5pt] (0,-4) -- (4,0) node[pos=.7, below right]{$I^-$};
	\draw [<->] (1.85, 2.45) -- (2.15, 2.15) node[above]{$\epsilon$};
	\draw [<->] (1.85, -2.45) -- (2.15, -2.15) node[below]{$\epsilon$};
	
	\draw[dotted] (0,0)	-- (2,2);
	\draw[dotted] (0,0)	-- (2,-2);
	
	\draw [dotted](0,3.4) ..controls (1,2.8).. (2,2);
	\draw [dotted](0,2.8) ..controls (.9,2.6).. (2,2);
	\draw [dotted](0,2) -- (2,2);
	\draw [dotted](0,1.3) ..controls (.9,1.4).. (2,2);
	\draw [dotted](0,.6) ..controls (1,1.2).. (2,2);

	\draw [dotted](0,-3.4) ..controls (1,-2.8).. (2,-2);
	\draw [dotted](0,-2.8) ..controls (.9,-2.6).. (2,-2);
	\draw [dotted](0,-2) -- (2,-2);
	\draw [dotted](0,-1.3) ..controls (.9,-1.4).. (2,-2);
	\draw [dotted](0,-.6) ..controls (1,-1.2).. (2,-2);

	\draw [dotted](2,2) -- (2,-2);
	\draw [dotted](2,2) ..controls (0,0).. (2,-2);
	\draw [dotted](2,2) ..controls (1,0).. (2,-2);
	\draw [dotted](2,2) ..controls (3,0).. (2,-2);
	\draw [dotted](2,2) ..controls (4,0).. (2,-2);
		\end{tikzpicture}
			\end{center}
		\end{minipage}
	\begin{minipage}{8cm}
		\begin{center}
					\begin{tikzpicture}[scale=1]		
	\shade [shading=axis,left color=white, right color=gray]  (4,0)--(2.5,-1.5) .. controls (3,0).. (2.5,1.5);
	\fill (0,4) circle (2pt) node[above] {$i^+$};
	\fill (0,-4) circle (2pt) node[below] {$i^-$};
	\fill (4,0) circle (2pt) node[right] {$i^0$};
	\draw [line width = 1.5pt, dashed] (0,4) -- (-1.5,2.5);
	\draw [line width = 1.5pt, dashed] (0,-4) -- (-1.2,-2.5);

	\draw [line width = 1.5pt, red] (2.5,-1.5) .. controls (3,0).. (2.5,1.5);

	\draw [line width = 1.5pt] (0,4) -- (4,0) node[pos=.5, above right]{$I^+$};
	\draw [line width = 1.5pt] (0,-4) -- (4,0) node[pos=.5, below right]{$I^-$};
	\draw [line width = 1.5pt, dashed] (0,4) -- (-1.5,4);
	\draw [line width = 1.5pt, dashed] (0,-4) -- (-1.5,-4);
	
	\draw (1, 0) -- (2.5,1.5);
	\draw (1, 0) -- (2.5,-1.5);
	\draw [dashed] (0,-4) .. controls (1.3,0).. (0,4) node[pos=.5, above, sloped]{$r=r_0$};
	\draw (3.4,0) node {$\mathcal{D}_{\omega}$};

		\end{tikzpicture}
		\end{center}
	\end{minipage}
	\caption{\emph{Left.} Dotted lines represent the level sets of $\frac{1}{(-u)v}$ in Minkowski spacetime. One level set of $f_{\epsilon} = \frac{1}{(-u + \epsilon)(v + \epsilon)}$ is also shown (red line). In order to ensure the unique continuation property for a wave in Minkowski spacetime one has to prescribe initial data (to all orders) on more than half of null infinity. \emph{Right.} One level set of $f= \frac{1}{(-u)v}$ is shown in Schwarzschild spacetime as well as the corresponding neighbourhood of infinity $\mathcal{D}_{\omega}$. The pseudo-convex function depends now on a parameter $r_0>2M$; different choices of $r_0$ give rise to `parallel' foliations. This behaviour is responsible for the localised result around spatial infinity for positive-mass spacetimes: Data require for unique continuation from infinity can be provided on small portions of null infinity.}
	\end{figure}
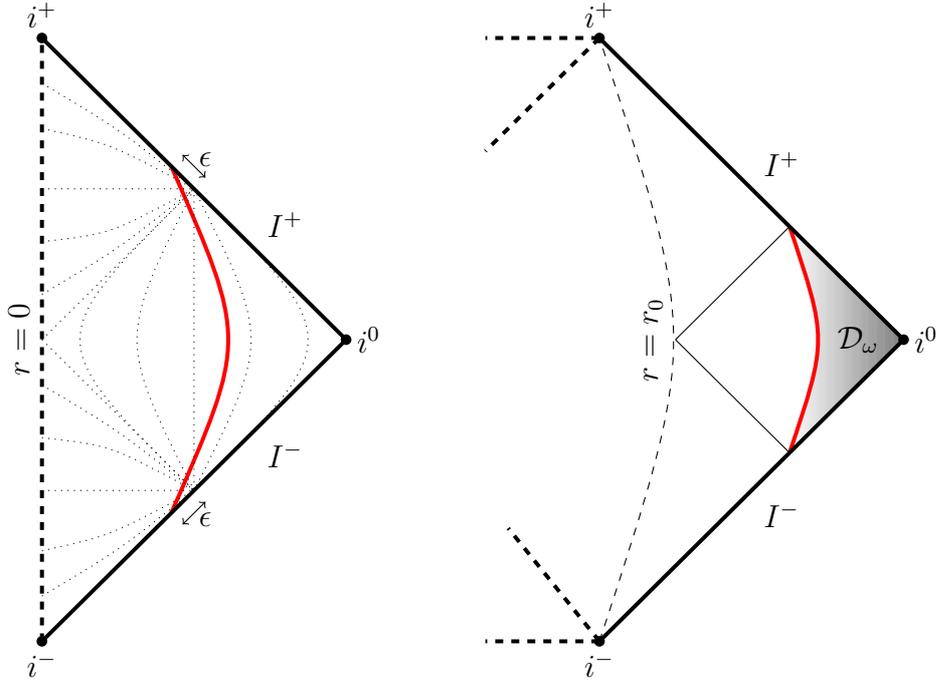

	Now, an important ingredient in the deduction of the Carleman estimates is a \emph{pseudo-convex} function. Geometrically, in the Lorentzian context, these are functions whose level sets are convex with respect to null geodesics; that is, any null geodesic tangent to a level set locally remains on one side of that level set. This is equivalent to the following quantitative condition, \cite{Hormander}:  A function $f$ on $(\mathcal{M},g)$ is pseudo-convex if there exists a function $h$ such that 
	\[	\pi := h g - \nabla^2 f \]
	is positive definite when restricted to the tangent space of the level sets of $f$. Alexakis-Schlue-Shao proved that, with respect to the above coordinates, the function
	$$  f:= \frac{1}{(-u)v},   	$$
	is pseudo-convex. Therefore we expect the unique continuation property to hold across its level sets. One technicality arises in that this pseudo-convexity degenerates towards infinity. To cope with this degeneracy, which takes the form of vanishing/blowing up weights towards infinity, Alexakis-Schlue-Shao rely on a reparametrisation of $f$, 
	$$ 	F(f) := \log f - f^{2\delta} ,$$
	for some $\delta>0$ to be chosen later. Then they are able to conclude the unique continuation from infinity property in a neighbourhood of the form 
	\[	\mathcal{D}_{\omega} := \{ (u,v,y^i) : 0 < f(u,v) < \omega	\}	. \]
	In order to understand the geometry of pseudo-convex timelike hypersurfaces we start by analysing them in Minkoswki spacetime. Consider double null coordinates $(u,v, \theta^2, \theta^3)$. The pseudo-convex timelike hypersurfaces considered by Alexakis-Schlue-Shao \cite{ASS} are given by the positive level sets of the function $f_{\epsilon} = \frac{1}{(-u + \epsilon)(v + \epsilon)}$, $\epsilon>0$. The $\epsilon$-perturbation is necessary to accomplish the pseudo-convexity condition in the absence of a mass. See Figure \ref{fig:pseudo-convex}. \\
	
	On positive-mass spacetimes the situation is qualitatively different. Consider for example Schwarzschild spacetime in double null coordinates, $(u,v,\theta^2, \theta^3)$, recall that these are defined by 
	$$	u = \frac{t-r_* }{2} , \qquad v = \frac{t+r_* }{2} , $$
	where 
	$$	r_*(r) = \int_{r_0}^r\left(	1- \frac{2M}{s}\right)^{-1}ds, \qquad	r_0>2M .	 $$
	Due to the presence of a mass now the function $f = \frac{1}{(-u )v}$ is pseudo-convex. \\

We are now in position to state the main technical tool of this section. For convenience we introduce the weight function $\mathcal{W}$ and associated weighted norms. For any $\la>0$ and domain $\mathcal{D}=\mathcal{D}_{\omega}$, $\omega>0$
	$$
	\mathcal{W} = \e^{-\lambda F} f^{\frac12}, \quad	\| \cdot \|_{\mathcal{W}} = \| \mathcal{W} \cdot \|_2, 	\quad	\| \phi \|^2_2 = \int_{\mathcal{D}} |\phi|^2 d\mu_g .
	$$	
	
	\begin{thm}	\label{thm:Carleman} \emph{(Carleman estimate near infinity for linear waves, \cite{AS}) } Let $(\mathcal{M},g)$ be an asymptotically flat spacetime with positive mass $m\geq m_{min}>0$ in the sense of (\ref{metric AS}) and $\mathcal{D}_{\omega}$ a neighbourhood of infinity for some $\omega>0$. Let $\delta>0$ and let $\phi$ be a smooth function on $\mathcal{D}_{\omega}$ that vanishes to all orders at infinity in the sense that for each $N\in \N$ there exist an exhaustion\footnote{A nested family of subsets, with piece-wise smooth time-like boundaries, whose union is all of $\mathcal{D}_{\omega}$.} $(\mathcal{U}_k)$ of $\mathcal{D}_{\omega}$ such that
	\begin{align} \label{vanishing condition}
	\lim_{k\rightarrow \infty}	\int_{\partial \mathcal{U}_k} r^N (\phi^2+|\partial \phi|^2) = 0 .	
	\end{align}
	Then, for $\omega>0$ sufficiently small and $\la>0$ sufficiently large, 
	\begin{align}	\label{Carleman}
		\la^3 \| f^\delta \phi	\|_{\mathcal{W}} + \la \| f^{-\frac12} \Psi^{\frac12} \nabla \phi  \|_{\mathcal{W}} \lesssim \| f^{-1} \square \phi \|_{\mathcal{W}} ,
	\end{align}
	where $\Psi$ is defined by 
		\[	\Psi := \frac{ m_{min} \log r}{r} .	\]
	\end{thm}

	\textbf{Sketch of proof.} Consider $\varphi = \e^{-\la F(f)} \phi$. The idea is to obtain an energy estimate for $\varphi$, but here we wish for the bulk terms of the integral to be positive and for the boundary terms to vanish. That is, consider the modified energy current
	\[ J_{\beta}^{w}[\varphi] = Q_{\al \be}[\varphi] \nabla^{\al} f + \left( \frac12 \partial_{\be} w + \frac12 \la^2 (\nabla^{\al}f)(\nabla_{\al}f)(F')^2 \partial_{\be} f \right)  \varphi^2 - \frac12 w\partial_{\be}(\varphi^2) , 	\]
	where $Q_{\al \be}[\varphi]$ is the standard energy-momentum tensor for $\square \varphi=0$. \\
	
	The function $w$ is to be chosen appropriately so that $\divergence J_{\beta}^{w}$ produces the tensor $\pi$, hence capturing the pseudo-convexity of $f$. Specifically, the choice
	$$ w = h - \frac12 \square f - \frac12, $$
	produces positive bulk terms that are quadratic in $\partial_X \varphi$ for $X$ tangent to the level sets of $f$; in order to obtain positivity in the normal direction one relies on the choice of reparametrisation $F(f)$. 	\\
	
	The above procedure ultimately results in an inequality of the form 
		\begin{align*}
				 \int_{\mathcal{D}_{\omega}}	\mathcal{W}_L |\mathcal{P} \varphi|^2 \geq 	& \, 
								C \la \int_{\mathcal{D}_{\omega}} \left( \mathcal{W}_N |\nabla_N \varphi|^2 +\mathcal{W}_T |\nabla_T \varphi|^2 +\mathcal{W}_T \sum |\nabla_{e_i} \varphi|^2 		\right)	\\
								& C \la^3 \int_{\mathcal{D}_{\omega}}	\mathcal{W}_0 |\varphi|^2 + \int_{\mathcal{D}_{\omega}}	\mathcal{E}, 
		\end{align*}
	where $\mathcal{P} (\varphi) = \e^{-\la F(f)}\square(\e^{\la F(f)} \varphi)$ and $\mathcal{W}_L$, $\mathcal{W}_N$, $\mathcal{W}_T$, $\mathcal{W}_0$ are positive weights. The only non-positive term is the error $\mathcal{E}$ which must be absorbed into the remaining positive terms. 
	 $_{\blacksquare}$\\

Alexakis-Schlue-Shao used the above Carleman estimate to extend the vanishing of a function $\phi$ into a neighbourhood of infinity, \cite{ASS}. The proof is based on standard arguments using cut-off functions to cope with the ``inner decay'' necessary for Theorem \ref{thm:Carleman} to hold. The wave equation is used to substitute $\square \phi$ by lower order derivatives which can be absorbed on the left-hand side of (\ref{Carleman}) due to the decaying conditions. Finally the result follows by taking $\la$ to infinity. We omit a detailed proof since we will follow the same program in the proof of Proposition \ref{prop:unique continuation2}.	\\

We will also need Carleman estimates for transport equations involving $\Lie_T g$ and its first derivatives. These are covered by the following lemma proved in \cite{AS}.

\begin{lem}	\label{lem:CarlemanODE}
Let $(\mathcal{M}, g)$ be an asymptotically flat spacetime and $L=\partial_v$ be the outgoing null vector field in coordinates (\ref{metric AS}). Let $\phi$ be a smooth function on $\mathcal{D}_{\omega}$ that vanishes to all orders at infinity, in the sense that for any $N\in \N$ there is an exhaustion $(\mathcal{U}_k)$ of $\mathcal{D}_{\omega}$ such that
	$$\lim_{k\rightarrow \infty} \int_{\partial \mathcal{U}_k} r^N\phi^2 =0	. $$
Then for any $q\geq 1$ and $\la>0$ sufficiently large,
	$$ \la \|\frac1r f^{-1}r^{-q} \phi \|_{\mathcal{W}} \lesssim \|f^{-1} r^{-q} \nabla_L \phi	\|_{\mathcal{W}} . $$
\end{lem}	

To conclude this section we present a technical result necessary to ensure that we can change from time-space coordinates in the asymptotically flat sense of definition \ref{def:AF} to the double-null coordinates required for the Carleman estimates. 

\begin{lem}\label{lem:doublenull}
	Let $(\mathcal{M},g, F)$ be the Cauchy development of initial data suitably close to Kerr-Newman. Specifically, let $g|_{\Sigma}$ and $\partial_t g|_{\Sigma}$ be induced on $\Sigma= \{ t=0	\}$ by 
	\begin{align}	\label{Kerr-Newman}
		g &=  g_{KN} + g^{\infty}  \nonumber 	\\
			& = - \left( 1- \frac{2M}{r} + \frac{e^2}{r^2} + \order^{\infty}_2(r^{-3})		\right) dt^2 
									+ \order^{\infty}_2(r^{-4}) dt dr  \nonumber \\
							& +	\left( 1 + \frac{2M}{r} - \frac{a^2 + e^2}{r^2}\sin^2\vartheta^2 + \frac{2Ma^2}{r^3}\cos^2\vartheta^2 + \order^{\infty}_2(r^{-4})		\right) dr^2 	\nonumber \\
							& + \left( \breve{\gamma}_{ij} + \order^{\infty}_2(1)	\right) d\vartheta^i d\vartheta^j + 
							\sum_{i=2}^3 \left(	 \order^{\infty}_2(r^{-1})dt d\vartheta^i	+ \order^{\infty}_2(r^{-3})dr d\vartheta^i \right)
	\end{align}		
	Suppose $(\mathcal{M},g)$ is asymptotically flat, in the sense of Definition \ref{def:AF}. Moreover, assume that the vector field $T$, as constructed in Section \ref{section:tetrad and coordinates}, is Killing to order 4, that is, 
			\[	\lim_{r \rightarrow \infty} r^N \Lie_T g = 0 	\qquad N=0,\ldots,4. \]
	Then we can change to the double-null coordinates required for the Carleman estimates of Theorem \ref{thm:Carleman}, i.e., there exists a domain $\mathcal{D}=(-\infty, 0)\times (0, \infty) \times \mathbb{S}^2 \subset \mathcal{M}$ with coordinates $(u,v, y^i)$  such that the metric components verify condition (\ref{metric AS}) and the corresponding bounds. 
\end{lem}
\textbf{Proof.} We will prove that the asymptotic expansion (\ref{Kerr-Newman}) holds through out the development when written in $T$-adapted coordinates. Then a transformation to `comoving coordinates', as in \cite{ASS}, shows that $(\mathcal{D}, g)$ is of the desired form. \\

First we need to find estimates for $T$ at $\Sigma$ with respect to the given coordinate system. Let us write,
	\[ T = \tau \partial_t + \rho  \partial_r + \theta^i \partial_{\vartheta^i} . \]
Recall that $T = \partial_u = \frac{L + \underline{L}}{2} + \sum_{\mu=0}^3 \order(r^{-1})e_\mu $. 	\\

We have abused notation by using the same symbol, $r$, for the area function of Section \ref{section:tetrad and coordinates} and the coordinate function in (\ref{Kerr-Newman}). However, due to the asymptotic expansion (\ref{Kerr-Newman}) assumed on $\Sigma$ we have that they agree up to $\order(1)$, hence the symbol $\order(r^{-k})$ is unambiguous. Moreover, the cone $C^-_{d^*}$ was chosen so that intersects $\Sigma$ on $r=d^*$, this guarantee that $L$ and $\underline{L}$ agree to leading order with the in-going and out-going null directions defined by the level sets of $r$. Therefore, given the expansion (\ref{Kerr-Newman}), $T$ must agree with $\partial_t$ to leading order, that is, $\tau \rightarrow 1$ and $\rho, \theta^i \rightarrow 0$ as $r \rightarrow \infty$ along $\Sigma$.\\

Now we use the hypothesis that $T$ is Killing to order 4 to find more refined estimates for  $\tau, \rho, \theta^i$ at $\Sigma$. Indeed, we have,
	\begin{align*}
		(\Lie_T g)_{\mu \nu}  & =  	(\partial_{\mu} \tau) g_{t \nu} + (\partial_{\nu} \tau) g_{\mu t} + \tau(\partial_t g)_{\mu \nu} \\
								& \quad +  	(\partial_{\mu} \rho) g_{r \nu} + (\partial_{\nu} \rho) g_{\mu r} + \rho(\partial_r g)_{\mu \nu} \\
								& \quad +	(\partial_{\mu} \theta^i) g_{\vartheta^i \nu} + (\partial_{\nu} \theta^i) g_{\mu \vartheta^i} + \theta^i(\partial_{\vartheta^i} g)_{\mu \nu}  .
	\end{align*}
Also, (\ref{Kerr-Newman}) gives control on $\partial g$ along $\Sigma$:
	\begin{align*}
		(\partial_t g)_{rr} &= \order^{\infty}_1(r^{-4}), &\quad (\partial_t g)_{rt} &= \order^{\infty}_1(r^{-4}), &\quad 
									(\partial_t g)_{r \vartheta^i} &= \order^{\infty}_1(r^{-3}), \\
		(\partial_r g)_{rr} &= -\frac{2M}{r^2} + \order^{\infty}_1(r^{-4}), &\quad (\partial_r g)_{rt} &= \order^{\infty}_1(r^{-5}), &\quad 
									(\partial_r g)_{r \vartheta^i} &= \order^{\infty}_1(r^{-4}), \\
		(\partial_{\vartheta^i} g)_{rr} &= \order^{\infty}_1(r^{-2}), &\quad (\partial_{\vartheta^i} g)_{rt} &= \order^{\infty}_1(r^{-4}), &\quad 				(\partial_{\vartheta^i} g)_{r \vartheta^j} &= \order^{\infty}_1(r^{-3}) .								
	\end{align*}
	Therefore, to leading order, the $(r,{\vartheta^i})$, $(r,r)$ and $(r,t)$ components of $\Lie_T g$ imply the following differential estimates:
	\begin{align*}
		\partial_r \theta^i + \frac{1}{r^2}(\partial_{\vartheta^i} \rho) + \frac{1}{r^2}(\partial_r \tau)\order(r^{-1}) &= \order(r^{-4}) , \\
		2(\partial_r \rho)\left( 1 +\frac{2M}{r}	\right) - \left( \frac{2M}{r^2} + \order(r^{-2}) \right) \rho +  \theta^i \order(r^{-2}) & = \order(r^{-4}) , \\
		-2(\partial_r \tau)\left( 1-\frac{2M}{r} \right) + (\partial_r \theta)\order(r^{-1}) + \theta^1\order(r^{-3}) & = \order(r^{-4}) .
	\end{align*}		
Now, the a priori estimates, $\tau \rightarrow 1$ and $\rho, \theta^i \rightarrow 0$, together with the above bounds show that $\partial_r\theta^i, \partial_{r} \rho, \partial_r \tau = \order(r^{-2})$. Thus, 
	\begin{align}
		\tau-1, \rho, \theta^i = \order (r^{-3}). 
	\end{align}
Finally, we define $T$-adapted coordinates by setting $\bar{t}=0$ on $\Sigma$ and $T(\bar{t})=1$, while coordinates $(r, y^2, y^3)$ are defined to be constant along $T$-lines with $(r, y^2, y^3)= (r, \vartheta^2, \vartheta^3)$ on $\Sigma$. The goal of the previous analysis was to achieve the correct asymptotic behaviour of the metric in this new coordinates, indeed, it is easy to check that
	\begin{align*}
		g  = & - \left( 1- \frac{2M}{r} + \frac{e^2}{r^2} + \order^{\infty}_2(r^{-3})		\right) d\bar{t}^2 
									+ \order^{\infty}_2(r^{-4}) d\bar{t} dr  \nonumber \\
							& +	\left( 1 + \frac{2M}{r} - \frac{a^2 + e^2}{r^2}\sin^2y^2 + \frac{2Ma^2}{r^3}\cos^2y^2 + \order^{\infty}_2(r^{-4})		\right) dr^2 	\nonumber \\
							& + \left( \breve{\gamma}_{ij} + \order^{\infty}_2(1)	\right) dy^i dy^j + 
							\sum_{i=2}^3 \left(	 \order^{\infty}_2(r^{-1})d\bar{t} dy^i	+ \order^{\infty}_2(r^{-3})dr dy^i \right)
	\end{align*}		
In these coordinates, the metric components are easily controlled by the initial data and $\Lie_T g$, indeed, 
	\[ \left| g_{\mu \nu} (\bar{t}, r, y^2, y^3) - g_{\mu \nu} (0,r, y^2, y^3) \right| \leq	
				\int_0^{\bar{t}} | (\Lie_T g)_{\mu \nu} (t,r,y^2, y^3)| dt .	\]
Therefore on a domain of the form,
	$$	\mathcal{D} = \{ (\bar{t}, r, y^2, y^3) : r\geq R, |\bar{t}| \leq r + 2M\log r - (R+2M\log R) \} ,	$$
we have that $|g_{\mu \nu} (\bar{t}) - g_{\mu \nu}(0)| = \order(r^{-3})$. \\

Then a transformation to comoving coordinates as in \cite{ASS} achieves the desired double-null form (\ref{metric AS}). $_{\blacksquare}$

\subsection{Proof of the main Theorem}

Now we proceed to the last part of the proof of Theorem \ref{thm1}. Namely, we show that the vector field $T=\partial_u$ is indeed a symmetry of the gravitational and electromagnetic fields provided that it is a symmetry to all orders at infinity. More precisely we have the following:

\begin{prop}	\label{prop:unique continuation2}
Let $(M,g,F)$ be an asymptotically flat electrovacuum spacetime with rapidly decaying curvature in the sense that uniformly with respect to an orthonormal frame
	\begin{align} \label{decay}
	\begin{aligned}
		|C_{\al \be \ga \de}| & = \order(r^{-3}),  \\	|\nabla_\ep C_{\al \be \ga \de}| &=\order(r^{-4}),	
	\end{aligned}
	&&
	\begin{aligned}
		|F_{\al \be}| &=\order(r^{-2}),  \\ 	|\nabla_{\ep} F_{\al \be}| &=\order(r^{-3}).
	\end{aligned}
	\end{align}
Let $(u,s, \theta^2,\theta^3)$ be coordinates on $D_{\omega}$, $\omega>0$, constructed as in Section \ref{section:tetrad and coordinates} with $L:=\partial_s$ and $T:=\partial_u$ satisfying 
	\[	\nabla_L L=0 , \qquad	[L,T]=0	. \]
Assume that $T$ is a symmetry to all orders at infinity, i.e., for all $N\in \N$ there is an exhaustion $(\mathcal{U}_k)$ of $D_{\omega}$ such that
	\begin{align}	\label{killing to all orders}
		\lim_{k\rightarrow \infty} \int_{\partial \mathcal{U}_k} r^N \Lie_T g = 0, 	\qquad
		\lim_{k\rightarrow \infty} \int_{\partial \mathcal{U}_k} r^N \Lie_T C = 0, 	\qquad
		\lim_{k\rightarrow \infty} \int_{\partial \mathcal{U}_k} r^N \Lie_T F = 0.
	\end{align}
Then $T$ is in fact locally a genuine symmetry for $(\mathcal{M},g,F)$, namely
	\[	\Lie_Tg \equiv 0, \quad \Lie_T C \equiv 0, \quad \Lie_T F \equiv 0, 	\quad \textrm{ on }	 \mathcal{D}_{\omega'}\]
	for some $0<\omega'<\omega$.
\end{prop}

Note that conditions (\ref{decay}) corresponds precisely to the non-radiating hypothesis of Theorem \ref{thm1}; while conditions (\ref{killing to all orders}) are those deduced from Corollary \ref{cor:non-radiating}. \\

The strategy to prove Proposition \ref{prop:unique continuation2} is analogous to the one used by Ionescu-Klainerman in \cite{IK} and Alexakis-Schlue in \cite{AS}. We start by deducing tensorial wave equations for $\Lie_T C$ and $\Lie_T F$. When written in Cartesian coordinates those equations fit into the framework provided by the Carleman estimates of Theorem \ref{thm:Carleman}. Once we get those differential estimates a standard argument to deduce unique continuation follows through.\\	

We emphasise here that the full system of equations, (\ref{coupled system}) below, does not fit into the Alexakis-Schlue argument since one of the coupling terms does not decay fast enough. We deal with this problem by borrowing decay from the well-behaved coefficients and by using different Carleman weights for each of $\Lie_T C$ and $\Lie_T F$. See equations (\ref{CarlemanW3}), (\ref{CarlemanE}) and ensuing discussion.

\subsection{Ionescu-Klainerman tensorial equations}

In this section we recall the equations satisfied by the deformation tensors deduced by Ionescu and Klainerman in \cite{IK}. We remind the reader that their work assumes a vacuum spacetime, that is, the Riemann and Weyl tensors are equal, $R_{abcd}=C_{abcd}$. This is in contrast with our current approach where the Riemann tensor is coupled to the Faraday tensor via the Einstein equations. Hence we need to be careful now to distinguish between $R$ and $C$.	\\
\\
\textit{Notation.}	Through out this section we will denote schematically by
	\[ U \cdot V	\]
any linear combination of the product of two tensors $U$ and $V$, and contractions thereof. For example the relation 
	\[	
	R_{abcd} = C_{abcd} + \frac{1}{n-1} \left( g_{ac} S_{bd} - g_{dc} S_{ba} + g_{db} S_{ca} - g_{ab} S_{cd} \right) .
	\]
	will be abbreviated to
	\[	R = C+ g \cdot S . \\ \]
	
We proceed now to compute the Ionescu-Klainerman tensorial wave equations for $\Lie_T C$ and $\Lie_T F$. The idea is to use the wave equations satisfied by $C$ and $F$ and then commute the covariant and Lie derivatives with the help of Lemma \ref{lem:commutator}. The relevant equations are naturally coupled to $\pi_{ab}:=\Lie_T g_{ab}$ and $\nabla_a \pi_{bc}$, hence the necessity of finding (transport) equations for them. In order to get a closed system of equations we need to perform some algebraic tricks (related to the symmetries of the Weyl tensor) and work instead with auxiliary variables, $B_{ab}$ and $P_{abc}$; see Proposition \ref{prop:tensorial equations}. This section is entirely based on \cite{IK}.\\

We begin by noticing that $\Lie_T C$ is not trace-less. Indeed,
	$$	g^{ac} \Lie_T C_{abcd} = (-\Lie_T g^{ac})C_{abcd} = \pi^{ac}C_{abcd} .	$$
To remedy this we define the modified Lie derivative
	\begin{align}
		\hat{\Lie}_T C := \Lie_T C - B\odot C .
	\end{align}
where $B$ is a 2-covariant tensor and $(B\odot C)_{abcd} := {B_a}^eC_{ebcd} + {B_b}^eC_{aecd}+ {B_c}^eC_{abed}+ {B_d}^eC_{abce}$. If we take
	$$B = \frac12( \pi + \omega) ,$$
with $\omega$ any antisymmetric 2-form, a simple calculation leads to

\begin{lem} 
	The modified Lie derivative of the Weyl tensor, $W:= \hat{\Lie}_T C$, is a Weyl field, i.e.,
	\begin{enumerate}[i)]
		\item	$W_{abcd} =-W_{bacd} = -W_{abdc} = W_{cdab}$. 
		\item $W_{abcd} + W_{acdb} + W_{abdc} = 0 $.
		\item $g^{ac}W_{abcd} = 0$.
	\end{enumerate}
\end{lem}

Next, we need to compute the commutator of the Lie and covariant derivatives. This is given by the following:

	\begin{lem} \label{lem:commutator}
	For an arbitrary $k$-covariant tensor $V$ and vector field $T$ we have,
		\begin{align}
			\nabla_b \Lie_T V_{a_1...a_k} - \Lie_T \nabla_b V_{a_1...a_k} &= \sum_{j=1}^k \Pi_{a_jb\rho} {V_{a_1...}}^{\rho}{}_{...a_k} ,	 \\
			\Pi_{abc} & := \frac12 \left( \nabla_a \pi_{bc} + \nabla_b \pi_{ca} - \nabla_c \pi_{ab}	\right) .		
		\end{align}
	Schematically we write, $[\nabla_b, \Lie_T] V = \Pi_b \odot V$.
	\end{lem}
	\textbf{Proof.} We compute
		\begin{align*}
			\Lie_T V_{a_1 \ldots a_n} &= T^c \nabla_c V_{a_1 \ldots a_n} + \sum_i (\nabla_{a_i}T^c)V_{a_1 \ldots c \ldots a_n} ,	\\
			\nabla_b \Lie_T V_{a_1 \ldots a_n} &= (\nabla_b T^c ) \nabla_c V_{a_1 \ldots a_n} + T^c\nabla_b \nabla_c V_{a_1 \ldots a_n} \\
				& \qquad + \sum_i (\nabla_b \nabla_{a_i}T^c)V_{a_1 \ldots c \ldots a_n} + (\nabla_{a_i}T^c) \nabla_b V_{a_1 \ldots c \ldots a_n}, \\
			\Lie_T \nabla_b V_{a_1 \ldots a_n} &= T^c \nabla_c \nabla_b V_{a_1 \ldots a_n} + (\nabla_b T^c) \nabla_c V_{a_1 \ldots a_n} +
			\sum_i (\nabla_{a_i}T^c) \nabla_b V_{a_1 \ldots c \ldots a_n} .
		\end{align*}
		Then
			\begin{align*}
			[\Lie_T, \nabla_b] V_{a_1 \ldots a_n} = \sum_i \left ( T^c {R_{cba_i}}^f V_{a_1 \ldots f \ldots a_n} +  
				(\nabla_b \nabla_{a_i} T^c) V_{a_1 \ldots c \ldots a_n} \right).
			\end{align*}
		And the results follows from the identity
			\begin{align*}
				\nabla_b\nabla_a T_c = R_{cabd}T^d + \Pi_{abc}.
			\end{align*}
		To prove this equality we just evaluate, commute derivatives and make use of the 1st Bianchi identity:
			\begin{align*}
				R_{cabd}T^d + \Pi_{abc} &= R_{cabd}T^d + \frac12 \left( \nabla_a (\nabla_b T_c + \nabla_cT_b) + \nabla_b(\nabla_aT_c + 				  											\nabla_cT_a) - \nabla_c(\nabla_aT_b + \nabla_bT_a)\right), \\
						&= R_{cabd}T^d + \frac12 \left( R_{acbd}T^d + R_{bcad}T^d + 2 \nabla_a\nabla_bT_c - R_{bacd}T^d \right), \\
						&= R_{cabd}T^d + \frac12 \left( 2R_{acbd}T^d + 2 \nabla_a\nabla_bT_c \right), \\
						&= \nabla_a\nabla_bT_c . \qquad {}_{\blacksquare}
			\end{align*}
	
	Now we present the variables and equations which play the crucial role in the unique continuation analysis. The variables are minor modifications of $\pi$, $\nabla \pi$ and $\Lie_T C$ which give us a closed system of equations.
	
	\begin{prop} \label{prop:tensorial equations}
	Let $\pi_{ab}:= \Lie_Tg_{ab}$ and $\omega_{ab}$ be a 2-form solution of the transport equation
		$$ \nabla_L \omega_{ab} = \pi_{ac} \nabla_b L^c - \pi_{bc} \nabla_a L^c .$$
	Define the tensors $B$, $P$ and $W$ as follows,
			\begin{align}
				B_{ab} &:= \frac12 \left( \pi_{ab} + \omega_{ab}	\right), 	\\
				P_{abc} &:= \frac12 \left( \nabla_a \pi_{bc} - \nabla_b \pi_{ac} -\nabla_c \omega_{ab} \right) = \Pi_{acb} - \nabla_c B_{ab}, \\
				W_{abcd} &:= \Lie_T C_{abcd} - (B\odot C)_{abcd}  ,
			\end{align}
	Then the following equations hold	
			\begin{align}
				\nabla_L B_{ab} &= L^c P_{cba} - B_{cb}\nabla_aL^c, \\
				\nabla_L P_{abc} & = L^d (\Lie_T R_{abcd} - {B_a}^e R_{ebcd}- {B_b}^e R_{aecd} + P_{abd} \nabla_cL^d, \\
				\square	W & = \Lie_T (\square C) + \nabla P \cdot C + B\cdot \nabla C + \nabla B \cdot \nabla C + B \cdot \square C.
			\end{align}
	\end{prop}
	
	During the proof we will make use of the following identities:
	
	\begin{lem}
	Assume $[T,L]=0$, $L^c\nabla_c L_a = 0$ and $\pi$ vanishes to first order at infinity. Then 
	$$L^b\pi_{ab}=0 , \quad P_{abc}L^c = 0, \quad	L^b \omega_{ab}=0 .$$
	\end{lem}
	\textbf{Proof of Lemma.} We start by showing $L^aL^b\pi_{ab}=0$. Indeed, 
		\begin{align*}
			L^aL^b\pi_{ab} &= L^aL^b (\nabla_a T_b + \nabla_b T_a ) = L^b T^a\nabla_a L_b + L^a T^b\nabla_b L_a	\\
							&= T^a \nabla_a (L_b L^b) = 0.
		\end{align*}		 
	Now, by commuting derivatives we can find a transport equation for $L^b\pi_{ab}$.
		\begin{align*}
			L^c\nabla_c (L^b\pi_{ab} ) &= L^c \nabla_c (L^b (\nabla_a T_b + \nabla_b T_a ) 	,\\
							&= L^c L^b ( \nabla_a \nabla_c T_b + {R_{cab}}^e T_e) + L^c \nabla_c (T^b \nabla_b L_a) ,	\\
							&= \nabla_a(L^cL^b) \nabla_c T_b + {R_{cab}}^e L^c L^b T_e) + {R_{cba}}^e L^cT^bL_e + T^c \nabla_c (L^b \nabla_b L_a),\\
							&= (\nabla_aL^c) (L^b \pi_{cb}) ,
		\end{align*}		 
	which is a homogeneous equation for $L^b \pi_{ab}$. In our context, the choice of $T$ implies that the deformation tensor $\pi$ vanishes to first order at infinity. In particular  $L^b\pi_{ab}$ also vanishes to first order at infinity; this together with the above transport equation implies that $L^b\pi_{ab}\equiv 0$ as desired.\\
	
	Next, a straightforward computation yields, 
	\begin{align*}
		2 P_{abc}L^c &= L^c(\nabla_a \pi_{bc} - \nabla_b \pi_{ac} -\nabla_c \omega_{ab})	, \\
						&= -\pi_{bc}\nabla_aL^c + \pi_{ca}\nabla_bL^c - \pi_{ac} \nabla_b L^c + \pi_{bc} \nabla_aL^c = 0, 
	\end{align*}
	where we have used $L^c\pi_{ac}=0$ and the definition of $\omega_{ab}$.\\

	Finally, for the last equality we deduce a transport equation for $L^b\omega_{ab}$ using its definition,
	\begin{align*}
		L^c \nabla_c(L^b\omega_{ab}) &= L^b L^c \nabla_c \omega_{ab} ,  \\
						&=  L^b (\pi_{ac} \nabla_bL^c - \pi_{bc} \nabla_aL^c) = 0 , 
	\end{align*}
	Hence $L^b\omega_{ab}$	is constant and we can choose initial conditions for $\omega_{ab}$ such that $L^b\omega_{ab}$ vanishes. $_{\blacksquare}$ \\
	
	\textbf{Proof of Proposition \ref{prop:tensorial equations}.} For the transport equation for $B_{ab}$ we compute:
		\begin{align*}
			2(L^c P_{cba} - B_{cb}\nabla_aL^c) &= L^c(\nabla_c\pi_{ba} - \nabla_b\pi_{ac} - \nabla_a\omega_{cb}) - (\pi_{cb}+\omega_{cb})\nabla_aL^c	\\
								&= \nabla_L \pi_{ab} + \pi_{ac}\nabla_b L^c + \omega_{cb}\nabla_aL^c - (\pi_{cb}+\omega_{cb})\nabla_aL^c	\\
								&= 2 \nabla_L B_{ab},
		\end{align*}
	where we have used $L^c\pi_{cb}=L^c\omega_{cb}=0$ and the transport equation defining $\omega_{ab}$.	\\
	
	Next we deduce the transport equation for $P_{abc}$. Recall the following identity\footnote{Which is basically an antisymmetrised identity for $\nabla_d \nabla_a (\nabla_{(b} T_{c)})$ necessary to cope with the symmetries of $R_{abcd}$. The idea to prove it is to commute derivatives.} proved in \cite{IK} for $\tilde{P}_{abc} := \frac12 (\nabla_a\pi_{bc} - \nabla_b\pi_{ac})$,
		$$	\nabla_d\tilde{P}_{abc} - \nabla_c\tilde{P}_{abd} = \Lie_T R_{abcd} - \frac12 {\pi_a}^f R_{fbcd} - \frac12 {\pi_b}^f R_{afcd}.	$$
	Note that $P_{abc} = \tilde{P}_{abc} - \frac12 \nabla_c \omega_{ab}$, hence
		\begin{align*}
			L^d(\nabla_d P_{abc} - \nabla_c P_{abd}) &= L^d \left(\Lie_T R_{abcd} - \frac12 {\pi_a}^f R_{fbcd} - \frac12 {\pi_b}^f R_{afcd} + \frac12 (\nabla_c\nabla_d \omega_{ab} - \nabla_d\nabla_c \omega_{ab}) \right),	\\
					&= L^d \left(\Lie_T R_{abcd} - \frac12 {\pi_a}^f R_{fbcd} - \frac12 {\pi_b}^f R_{afcd}  - \frac12 {\omega_a}^f R_{fbcd} - \frac12 {\omega_b}^f R_{afcd}  \right),	\\
					&= L^d \left(\Lie_T R_{abcd} - {B_a}^f R_{fbcd} - {B_b}^f R_{afcd}  \right),
		\end{align*}
	the final result follows by noticing that $L^d\nabla_c P_{abd} = - P_{abd}\nabla_cL^d$ since $L^dP_{abd}=0$.\\
	
	Finally, we deduce the wave equation for $W$ by commuting Lie and covariant derivatives. We are interested only in the general structure of the equations, in particular, in the coefficients accompanying $W$ and $\nabla W$. Hence we do not any longer keep track of the different contractions but just on the bilinear structure of products and the different terms involving our variables $P$, $B$ and $W$. \\
\\	
\textit{Notation.}	During the following computations we will substitute freely $B$ instead of $\pi$ since $\pi_{ab} = B_{ab} + B_{ba}$. \\

	We start now with a divergence equation for $W$.
	
	\begin{lem} The following holds:
		\begin{align*}
			 \nabla^a W_{abcd} = B^{ae}\nabla_e C_{abcd} + g^{af}({P_f}^e{}_a C_{ebcd} + {P_b}^e{}_f C_{aecd} + {P_c}^e{}_f C_{abed} + {P_d}^e{}_f C_{abce} ). 
		 \end{align*}
	\end{lem}
	\textbf{Proof.} We will prove the schematic version:
		$$ \divergence W = B \cdot \nabla C + P \cdot C  . $$
	We have that 
		\begin{align*}
			 \nabla_e W = \Lie_T (\nabla_e C) + \Pi_e \odot C - \nabla_e (B \odot C), 		 
		 \end{align*}
	hence, 
		\begin{align*}
			 \divergence W &= (\Lie_T g) \cdot \nabla C + \Lie_T \divergence C + (\Pi_e - \nabla_e B) \cdot C - B\cdot \nabla C, 		 \\
			 		&= B \cdot \nabla C + P \cdot C ,
		 \end{align*}
	where we used $\pi_{ab} = B_{ab} + B_{ba}$, $\Pi_{aeb} - \nabla_e B_{ab} = P_{abe}$ and $\divergence C = 0$. $_\blacksquare$	\\
	
	We proceed similarly to obtain the wave equation for $W$,
		\begin{align*}
			 \square W &= \nabla ^e( \Lie_T \nabla_e C + \Pi_e \odot C - \nabla_e(B \odot C) )  , 		 \\
			 		&=  \Lie_T \square C + \Pi^e \cdot \nabla_e C + \nabla ^e(\Pi_e \odot C - \nabla_e(B \odot C))  , 	\\
			 		&= \Lie_T \square C + \nabla B \cdot \nabla C + \nabla^e( \Pi_e - \nabla_e B) \odot C +  B\cdot \square C ,	\\
			 		&= \Lie_T \square C + \nabla B \cdot \nabla C + \nabla P \cdot C + B\cdot \square C,
		 \end{align*}
		 where we have used once more $\Pi_{aeb} - \nabla_e B_{ab} = P_{abe}$ and $\Pi=\nabla \pi = \nabla B$ (schematically). This finishes the proof of Proposition \ref{prop:tensorial equations}. $_{\blacksquare}$	\\
	
	Using the same argument as above we can prove the following general statement:
	
	\begin{lem}
	Let $F$ be a $k$-covariant tensor, then $E:= \Lie_T F- B\odot F $ obeys
		\begin{align}
			\square E = \Lie_T \square F + \nabla B \cdot \nabla F + \nabla P \cdot F + B\cdot \square F.
		\end{align}
	\end{lem}

	This last lemma can be applied to the Faraday tensor. To conclude this section we state the full system of equations relevant for the Carleman estimates. We start by recalling the wave equations satisfied by the Weyl and Faraday tensors, as well as the Einstein equations in schematic form: 
	\begin{align}
		\square C &= R \cdot C = (C + g\cdot S) \cdot C, \\
		\square F &= R \cdot F = (C + g\cdot S) \cdot F, \\
		S &= F \cdot F
	\end{align}	
	Hence 
		\begin{align*}
			\Lie_T\square C & = (\Lie_T C + \pi \cdot F^2 + \Lie_T F \cdot F)\cdot C + R \cdot \Lie_T C , \\
							& = (W + B\cdot C + B \cdot F^2 + E \cdot F + B \cdot F^2)\cdot C + R \cdot W + R\cdot B \cdot W.
		\end{align*}
		Similarly, for the modified Lie derivative of the Faraday tensor we obtain
		$$\Lie_T\square E  = (W + B\cdot C + B \cdot F^2 + E \cdot F + B \cdot F^2)\cdot F + R \cdot E + R\cdot B \cdot F . $$
		Therefore, we have proved:
	\begin{lem}
	The deformation tensors $W = \Lie_T C - B \odot C $ and $E = \Lie_T F - B\odot F$ satisfy the following wave equations
		\begin{align}
		\square W &= (R+C) \cdot W + (C^2 + F^2\cdot C + R\cdot C)\cdot B \nonumber 	\\
				& \quad + (F\cdot C) \cdot E + \nabla C \cdot \nabla B + C\cdot \nabla P, \\
		\square E &= F \cdot W + (F^3+ F\cdot C + R\cdot F)\cdot B \nonumber	\\
				& \quad + (F^2 + R)\cdot E + \nabla F\cdot \nabla B + F \cdot \nabla P .
		\end{align}		
	\end{lem}
	
\subsection{Estimates for the deformation tensors}

Now we use the Carleman estimates of Theorem \ref{thm:Carleman} applied to the wave equations for $W$ and $E$. We start by writing them in Cartesian coordinates. This type of coordinates are chosen due to its uniform decay in all directions. We also have to check that the fields $W$ and $E$ satisfy the vanishing condition (\ref{vanishing condition}). This latter condition is fulfilled on the ``exterior'' part thanks to the hypothesis of Proposition \ref{prop:unique continuation2}, that is, the decaying properties corresponds to the fields vanishing to all orders at infinity. However in order to cope with the ``interior'' decay also included in the vanishing condition (\ref{vanishing condition}) a cut-off function needs to be introduced.  Finally, by absorbing the lower order terms into the terms already present in the Carleman estimates we can conclude the vanishing of $W$, $E$, $B$ and $P$ in a neighbourhood of spatial infinity. 	\\

Firstly, we pass to Cartesian coordinates $(x^0, x^1, x^2, x^3)$ such that the metric takes the form 
	$$	g = -(dx^0)^2 + (dx^1)^2 + (dx^2)^2 + (dx^3)^2 + \sum_{\mu, \nu=0}^3 \order_2(r^{-1}) dx^{\mu}dx^{\nu} ,	$$
and $\partial_{x^0}|_{x^1, x^i}$ coincides with $\partial_u|_{s,\theta^i}$ as $r\rightarrow \infty$, where
	$$r = \sqrt{(x^1)^2 + (x^2)^2 + (x^3)^2}. $$
Then, in these coordinates the Christoffel symbols decay suitably fast. That is, 
	$$	\Gamma_{\mu \nu}^{\al} = \order_1(r^{-2}) \quad \textrm{and}	\quad \partial_{\be} \Gamma_{\mu \nu}^{\al} = \order(r^{-3})  .$$
In what follows, for brevity, we will denote by $(V) $ the components of a tensor $V$  with respect to these Cartesian coordinates. In particular, we have schematically
	\begin{align*}
		(\nabla V) &= \partial (V) + \Gamma \cdot (V) , \\
		(\square V) &= \square (V) + \Gamma \cdot \partial (V) + \partial \Gamma \cdot (V).
	\end{align*}
Hence, the wave equations for the components of $W$ and $E$ are
	\begin{align*}
		\square (W) & = [(R) + \partial \Gamma]\cdot (W) + \Gamma \cdot \partial (W) + [(F)\cdot (C)] \cdot  (E) + [(R)\cdot (C) + \partial (C) \cdot \Gamma]\cdot  (B) 		\\
		& \quad + \partial (C) \cdot  \partial (B) +  (C) \cdot  \partial (P) + (C)\cdot \Gamma \cdot  (P) , \\
		\square (E) & = [(F) + \partial \Gamma] \cdot  (W) +  \Gamma \cdot  \partial (E) + [(F)^2 + \partial\Gamma] \cdot (E) + [(R)\cdot (F) + \partial (F) \cdot \Gamma] \cdot  (B) 	\\ 
		& \quad + \partial (F) \cdot \partial (B) +  (F) \cdot \partial (P) + (F) \cdot \Gamma \cdot (P) ,
	\end{align*}
where we have just kept the leading order terms multiplying $W$, $E$, etc. In view of the asymptotic behaviour assumed for the Weyl and Faraday tensors we have the following estimates
	\begin{align} \label{coupled system}
		\begin{aligned}
		\square (W) & = \order(r^{-3}) (W) + \order(r^{-2})\partial (W) + \order(r^{-5}) (E) + \order(r^{-6}) (B) 		\\
		& \qquad + \order(r^{-4}) \partial (B) + \order(r^{-5}) (P) + \order(r^{-3}) \partial (P)  , \\
		\square (E) & = \order(r^{-2}) (W) +  \order(r^{-3}) (E) + \order(r^{-2})\partial (E) + \order(r^{-5}) (B) 	\\ 
		& \qquad + \order(r^{-3}) \partial (B) + \order(r^{-4}) (P)  +  \order(r^{-2}) \partial (P) .
		\end{aligned}
	\end{align}
While these are morally the reason for the unique continuation we still need to compensate for the fact that the coefficient accompanying $(W)$ in the second equation does not decay fast enough\footnote{We need a power strictly greater than 2 in order to run the Alexakis-Schlue argument, see proof below where the corresponding term is controlled by choosing different $\lambda$-weights for each Carleman inequality.}. However the coupling term in the first equation, $\order(r^{-5}) (E)$, allows us to borrow some decay by modifying the Carleman weight.\\

Before applying the Carleman estimates of Theorem \ref{thm:Carleman} we have to guarantee that all the quantities vanish to all orders in the sense of (\ref{vanishing condition}),
	$$	\lim_{k\rightarrow \infty } \int_{\partial \mathcal{U}_k} r^N(\phi^2 + |\partial\phi|^2) = 0.	$$
We deal with the ``interior'' and ``exterior'' parts of the the boundary of $\mathcal{U}_k$ differently. \\

The ``exterior'' boundary approaches infinity as $r\rightarrow \infty$ and so it captures the idea of $\phi$ vanishing to all orders at infinity. By assumption the quantities $\pi =\Lie_T g$, $\Lie_T C$ and $\Lie_T F$ vanish to all orders at infinity. Now we check the modified versions. Firstly, $(\omega)$ satisfies the transport equation
	$$	\nabla_L (\omega) = \order(r^{-1})(\pi). $$
with $(\omega)=0$ at infinity by construction. Thus, $(\omega)$ vanishes to all orders at infinity. It follows that 
	\begin{align*}
		(B) &= (\pi) + (\omega), \\
		(P) &= \partial (\pi + \omega) + \Gamma \cdot (\pi + \omega), \\
		(W) &= (\Lie_T C) + (R) \odot (B), \\
		(E) &= (\Lie_T F) + (F) \odot (B), 
	\end{align*}
also vanish to all orders at infinity.\\

To deal with the ``interior'' part of $\partial \mathcal{U}_k$ a cut-off function is used. This technique is standard for unique continuation problems. \\

Let $\chi$ be a cut-off  function whose level sets coincide with those of $F\circ f$, 
	$$ \chi =1 \, \textrm{on } \mathcal{D}_{\omega_0}, \qquad	\chi=0 \, \textrm{on } \mathcal{M}\setminus \mathcal{D}_{\omega_1}, \qquad \omega_0<\omega_1<\omega .	$$
Then the functions $\chi \cdot (W)$, $\chi \cdot (E)$, etc., satisfy the vanishing condition (\ref{vanishing condition}). The price to pay is that we have introduced extra terms in the wave equations, however these are easy to treat since they are supported only in the cut-off region. Indeed, 
	\begin{align*}
		\square (\chi\cdot (W)) &= (\square \chi) \cdot (W) + (\partial \chi) \cdot \partial (W) + \chi \cdot \square (W) , \\
								&= (\square \chi) \cdot (W) + (\partial \chi) \cdot \partial (W) + (\partial \chi) \cdot \{(W), \partial(W), \ldots \} 			\nonumber \\ 	& \quad+	\{ \chi \cdot (W), \partial (\chi \cdot (W)), \ldots  \} , \\
								&= \nabla \chi \mathbf{M} + \{ \chi \cdot (W), \partial (\chi \cdot (W)), \ldots  \} .
	\end{align*}
Hereafter we will use the symbol $\nabla \chi \mathbf{M} $ to denote multiples of $(W)$, $(E)$, etc., which are only supported in the cut-off region $\mathcal{D}_{\omega_1} \setminus \mathcal{D}_{\omega_0}$. Recall also that we have used the notation $\{ (W), (E), \ldots \}$ to denote a function involving $(W)$, $(E)$, etc. Hence, after applying the Carleman estimates we can focus only on the terms supported on $\mathcal{D}_{\omega_0}$.\\
\\
\textbf{Weighted Carleman estimates.}\\
\\
We are now ready to apply Theorem \ref{thm:Carleman} to the functions $\chi \cdot (W)$, $\chi \cdot (E)$, etc. To keep the notation simple and readable we will omit in the next argument the cut-off function and the parenthesis denoting Cartesian components. We follow the standard procedure to bound the $L^2$-norms of $W$, $B$, $P$ and its first derivatives. \\

The Carleman estimate for $W$ combined with its wave equation reads
	\begin{align}	\label{CarlemanW}
		\la^3 \| f^{\delta} W \|_{\mathcal{W}} + 	\la \| f^{-\frac12} \Psi^{\frac12} \partial W \|_{\mathcal{W}}
			 & \lesssim 	\| f^{-1} \square W \|_{\mathcal{W}}	, \nonumber \\
			& \lesssim \| f^{-1} r^{-3} W \|_{\mathcal{W}} + \| f^{-1} r^{-2} \partial W \|_{\mathcal{W}}  \nonumber	\\
			& \quad + \| f^{-1} r^{-6} B \|_{\mathcal{W}} + \| f^{-1} r^{-4} \partial B \|_{\mathcal{W}} \nonumber \\ 
			& \quad  + \| f^{-1} r^{-5} P \|_{\mathcal{W}}  + \| f^{-1} r^{-3} \partial P \|_{\mathcal{W}} 	\nonumber	\\
			& \quad + \| f^{-1} r^{-5} E \|_{\mathcal{W}} + \| \nabla\chi \textbf{M} \|_{\mathcal{W}} .
	\end{align}
The following estimates will be used throughout; they follow from the definition of $f$ and $\Psi$:
	\begin{align} \label{estimates f}
		f \gtrsim \frac{1}{r^2}, \qquad f\Psi \gtrsim \frac{1}{r^3}. 
	\end{align}	 
They imply
	\begin{align}
		 f^{-\frac12} \Psi^{\frac12} & =f^{-1}(f\Psi)^{\frac12} \gtrsim f^{-1}r^{-\frac32}, \\ \label{estimates f2}
		 f^{\delta}  &\gtrsim \frac{1}{r^{2\delta}} > \frac{1}{r},	\qquad \textrm{for } 0<2\delta < 1 .
	\end{align}
These last inequalities tell us that the first two terms on the right hand side of (\ref{CarlemanW}) can be absorbed into the corresponding terms on the left hand side, since $f^{-1}r^{-3} \lesssim f^{\delta}$ and $f^{-1}r^{-2} \lesssim f^{-\frac12} \Psi^{\frac12}$. This will be the main trick during the proof. \\

Next, we aim at controlling the $B$ terms. The Carleman estimate from Lemma (\ref{lem:CarlemanODE}) together with the transport equation for $B$ read
	\begin{align*}
		\la \|\frac1r f^{-1} r^{-4} B \|_{\mathcal{W}}  & \lesssim 	\| f^{-1} r^{-4} \nabla_L B \|_{\mathcal{W}}	, \nonumber \\
			& \lesssim \| f^{-1} r^{-4} P \|_{\mathcal{W}} + \| f^{-1} r^{-5} B \|_{\mathcal{W}} +  \| \nabla\chi \textbf{M} \|_{\mathcal{W}} .
	\end{align*}
We add this inequality to (\ref{CarlemanW}) and observe that the terms $\| f^{-1} r^{-5} B \|_{\mathcal{W}}$ and $\| f^{-1} r^{-6} B \|_{\mathcal{W}}$ can be absorbed into the left hand side. We have thus obtained,
	\begin{align}	 \label{CarlemanW2}
		\la^3 \| f^{\delta} W \|_{\mathcal{W}} + \la \| f^{-\frac12} \Psi^{\frac12} \partial W \|_{\mathcal{W}} + \la\| f^{-1} r^{-5} B \|_{\mathcal{W}} 
			& \lesssim \| f^{-1} r^{-5} E \|_{\mathcal{W}} + \| f^{-1} r^{-4} \partial B \|_{\mathcal{W}} \nonumber \\
			& \quad + \| f^{-1} r^{-4} P \|_{\mathcal{W}}  + \| f^{-1} r^{-3} \partial P \|_{\mathcal{W}} 	\nonumber	\\
			& \quad + \| \nabla\chi \textbf{M} \|_{\mathcal{W}} .
	\end{align}
Now we proceed similarly to bound $P$, $\partial B$ and $\partial P$. We have 
	\begin{align*}
		\la \|\frac1r & f^{-1} r^{-3} P \|_{\mathcal{W}}   \lesssim 	\| f^{-1} r^{-3} \nabla_L P \|_{\mathcal{W}}	, \nonumber \\
			& \lesssim \| f^{-1} r^{-3} W \|_{\mathcal{W}} + \| f^{-1} r^{-6} B \|_{\mathcal{W}} + \| f^{-1} r^{-5} E \|_{\mathcal{W}}  + \| f^{-1} r^{-4} P \|_{\mathcal{W}} + \|\nabla\chi \textbf{M} \|_{\mathcal{W}} , \\
		\la \|\frac1r & f^{-1} r^{-3} \partial B \|_{\mathcal{W}}  \lesssim 	\| f^{-1} r^{-3} \nabla_L \partial B \|_{\mathcal{W}}	, \nonumber \\
			& \lesssim \| f^{-1} r^{-5} B \|_{\mathcal{W}} + \| f^{-1} r^{-4} \partial B \|_{\mathcal{W}} + \| f^{-1} r^{-4} P \|_{\mathcal{W}} + \| f^{-1} r^{-3} \partial P \|_{\mathcal{W}} + \|\nabla\chi \textbf{M} \|_{\mathcal{W}} , \\
		\la \|\frac1r &f^{-1} r^{-2} \partial P \|_{\mathcal{W}}   \lesssim 	\| f^{-1} r^{-2} \nabla_L \partial P \|_{\mathcal{W}}	, \nonumber \\
			& \lesssim \| f^{-1} r^{-3} W \|_{\mathcal{W}} + \| f^{-1} r^{-2} \partial W \|_{\mathcal{W}} + \| f^{-1} r^{-6} B \|_{\mathcal{W}} + \| f^{-1} r^{-5} \partial B \|_{\mathcal{W}} \\
			& \qquad + \| f^{-1} r^{-4} P \|_{\mathcal{W}}  + \| f^{-1} r^{-5} E \|_{\mathcal{W}} + \| f^{-1} r^{-4} \partial E \|_{\mathcal{W}} + \|\nabla\chi \textbf{M} \|_{\mathcal{W}} .			
	\end{align*}
We add these inequalities to (\ref{CarlemanW2}) and observe that the $W$, $B$, $\partial B$, $P$ and $\partial P$ terms can be absorbed into the left hand side, thus obtaining,
	\begin{align}	 \label{CarlemanW3}
		\la^3 \| &f^{\delta} W \|_{\mathcal{W}} + \la \| f^{-\frac12} \Psi^{\frac12} \partial W \|_{\mathcal{W}} + \la\| f^{-1} r^{-5} B \|_{\mathcal{W}} + \la\| f^{-1} r^{-4} \partial B \|_{\mathcal{W}} 	\nonumber \\
		& + \la\| f^{-1} r^{-4} P \|_{\mathcal{W}} + \la\| f^{-1} r^{-3} \partial P \|_{\mathcal{W}} 
			 \lesssim \| f^{-1} r^{-5} E \|_{\mathcal{W}} +  \| \nabla\chi \textbf{M} \|_{\mathcal{W}} .
	\end{align}
for sufficiently large $\la$ and $0<\delta<\frac12$. An analogous argument gives (note the different Carleman parameter),
	\begin{align}	 \label{CarlemanE}
		\la'^3 \| &f^{\delta} E \|_{\mathcal{W}'} + \la' \| f^{-\frac12} \Psi^{\frac12} \partial E \|_{\mathcal{W}'} + \la' \| f^{-1} r^{-5} B \|_{\mathcal{W}'} + \la' \| f^{-1} r^{-4} \partial B \|_{\mathcal{W}'} 	\nonumber \\
		& + \la' \| f^{-1} r^{-3} P \|_{\mathcal{W}'} + \la' \| f^{-1} r^{-2} \partial P \|_{\mathcal{W}'} 
			 \lesssim \| f^{-1} r^{-2} W \|_{\mathcal{W}'} +  \| \nabla\chi \textbf{M} \|_{\mathcal{W}'} .
	\end{align}
We would like to add these last two inequalities and absorb the $W$ term on the left hand side to obtain the desired bound. However, for $\la =\la'$, this is not possible as $f^{-1} r^{-2}=\order(1)$ does not decay fast enough. To remedy this, we make the observation that the norms depend on $\la$ and by taking slightly different weights we can perform the procedure just described. More precisely, we want to find $\la'$ such that
	\begin{align} \label{inequalities}
		\begin{aligned}
		(e^{-\la F}f^{\frac12})f^{-1} r^{-5} & \lesssim e^{-\la' F} f^{\frac12} f^{\delta}, \\
		(e^{-\la'  F}f^{\frac12})f^{-1} r^{-2} & \lesssim e^{-\la F} f^{\frac12} f^{\delta}.
		\end{aligned}
	\end{align}
Indeed, we will show that the choice $\la' := \la - \delta$ achieves the previous inequalities. Firstly, note that
	\begin{align*}
		r^{-5} & \lesssim f^{1+2\delta}, \\
		r^{-2} & \lesssim f, 
	\end{align*}
these are a consequence of estimates (\ref{estimates f}) and (\ref{estimates f2}). They imply that 
	\begin{align*} 
		f^{-1} r^{-5} & \lesssim e^{\delta F}  f^{\delta}, \\
		e^{\delta F} f^{-1} r^{-2} & \lesssim f^{\delta},
	\end{align*}
since $F=\order(\log f) $. Finally, it is easy to see that these last inequalities are equivalent to (\ref{inequalities}), with $\la' := \la - \delta$. \\
\\
\textbf{Remark.} It is worth noticing that the previous argument did not make any special use of the power $r^{-5}$ accompanying $E$. The procedure will work for any $r^{-q}$ with $q>2$ by choosing $\delta>0$ small enough. \\

Now we are in position to close the argument. We add inequalities (\ref{CarlemanW3}) and (\ref{CarlemanE}), with $\la'= \la-\delta$. Inequalities (\ref{inequalities}) ensure that the terms $\| f^{-1} r^{-5} E \|_{\mathcal{W}}$ and $\| f^{-1} r^{-2} W \|_{\mathcal{W}'}$ can be absorbed into the left hand side. Moreover, the Carleman weights $e^{-\la F}$ and $e^{-\la' F}$ are monotonic increasing functions, so we can 
substitute its minimum value on left hand side and its maximum value on the right hand side (since these terms are supported in the cut-off region). Thus, after dropping the weight factors from the inequality we obtain the desired $L^2$-bound, 
	\begin{align*}	
		\la^3 \| f^{\frac12} f^{\delta} W \|_2 &+ \la \| \Psi^{\frac12} \partial W \|_2 + \la\| f^{\frac12} f^{-1} r^{-5} B \|_2 + \la\| f^{\frac12} f^{-1} r^{-4} \partial B \|_2	 \\
		& + \la\| f^{\frac12} f^{-1} r^{-4} P \|_2 + \la\| f^{\frac12} f^{-1} r^{-3} \partial P \|_2		\\
		& +	\la'^3 \| f^{\frac12} f^{\delta} E \|_2 + \la' \| \Psi^{\frac12} \partial E \|_2  + \la' \| f^{\frac12} f^{-1} r^{-5} B \|_2 + \la' \| f^{\frac12} f^{-1} r^{-4} \partial B \|_2 	 \\
		& + \la' \| f^{\frac12} f^{-1} r^{-3} P \|_2 + \la' \| f^{\frac12} f^{-1} r^{-2} \partial P \|_2  \\
		& \qquad		 \lesssim  \| \nabla\chi \textbf{M} \|_2 + \|  \nabla\chi \textbf{M} \|_2.
	\end{align*}
Finally, the left hand side can be interpreted as integrated over the smaller domain $\mathcal{D}_{\omega_0}$ where $\chi=1$ and by taking $\la \rightarrow \infty$ we conclude that $B \equiv 0$, $P \equiv 0$, $W\equiv 0$ and $E\equiv 0$ on $\mathcal{D}_{\omega_0}$. In particular
	$$	\Lie_T g \equiv 0 \qquad \textrm{and} \qquad \Lie_T F \equiv 0 \qquad \textrm{on } \mathcal{D}_{\omega_0} .$$
This finishes the proof of Proposition \ref{prop:unique continuation2} and Theorem \ref{thm1}. $_{\blacksquare}$

\appendix

\section{Einstein-Klein-Gordon system}
\setcounter{footnote}{1}

The goal of this work was to investigate the inheritance of symmetry property for the Einstein equations when matter/energy models are included. Indeed, we showed the validity of this property in the context of asymptotically flat electrovacuum spacetimes: An asymptotic time-like symmetry to all orders at infinity is indeed a (local) symmetry of both gravity and electromagnetism, Proposition \ref{prop:unique continuation2}. We also provided weaker conditions for the first condition to hold, namely, an asymptotic time-like symmetry to first order in a  non-radiating spacetime must be an asymptotic time-like symmetry to all orders, Corollary \ref{cor:non-radiating}. We also sketch the proof for the same results when a massless Klein-Gordon field is also present, but we stress that the conclusion no longer holds for the positive-mass case. \\

The assumed regularity assumptions are still a posteriori conditions, nevertheless, there are examples of spacetimes where our results apply. Moreover, the techniques employed during the proof are robust enough in that they can be extended to more general asymptotic expansions. However, the precise class of regularity conditions compatible with physical systems are still not well understood and a generalisation in this direction seems to need a different approach. Also, it is important to remark that the regularity assumptions used in this paper can be deduced from regular initial data if the resulting development is time-periodic, provided the preservation of regularity property at spatial infinity holds.	\\

	Now we proceed to discuss the conclusion of Theorem \ref{thm1} when other matter/energy models are considered. We present the result when a massless Klein-Gordon field is included; for simplicity we omit the Maxwell field in the following statements). The proof is entirely analogous to the one given in Sections 3 and 4. The conclusion is that the inheritance of symmetry holds as well for a massless Klein-Gordon field coupled to gravity. As remarked in the Introduction, this conclusion fails for massive matter fields, \cite{Bizon}, \cite{Chodosh}.\\
	
	We start by putting the result into context. In \cite{Dafermos}, Dafermos establishes a similar rigidity theorem for spherically symmetric Einstein-matter systems which are time-periodic. This corresponds, roughly speaking, to a ``no-hair'' result for spherically symmetric time-periodic black holes, thus generalising the work of Bekenstein \cite{Bekenstein}. More precisely, he concludes that solutions to asymptotically flat spherically symmetric time-periodic Einstein-matter systems are either Schwarzschild or Reissner-Nordstr{\"o}m spacetimes with vanishing matter fields. He assumes certain structure for the matter fields which includes, as examples, a wave map and a massive charged scalar field interacting with electromagnetism.	Also, another important assumption on the underlying spacetime is that of the existence a bifurcate horizon. Indeed, Dafermos' analysis (unlike ours) take place at the event horizon, where he shows vanishing of initial conditions for a 2D-characteristic problem which implies the vanishing of the fields in the domain of outer communications\footnote{In (1+1)-spacetime dimensions one has the nice property that an initial value problem set at a bifurcate horizon (or timelike hypersurface) is locally well-posed. This is seen by redefining the metric to be its negative.}. \\
	
	Here we adopt the ``far-away'' point of view and prove the following:
	\begin{thm}\label{thm:Klein-Gordon}
	\emph{(Stationarity of a non-radiating Einstein-massless-Klein-Gordon system.)} Let $(M,g, \phi)$ be an asymptotically flat non-radiating solution of the Einstein-massless-Klein-Gordon equations. Then there exists a timelike vector field $T$ such that in a neighbourhood of spatial infinity. 
			\[	\Lie_T g= 0 = \Lie_T \phi . 	\]
	\end{thm}	 
	
	Following the same procedure as in the proof of Theorem \ref{thm1} we split the proof in two parts. First, using the asymptotic expansion given by the asymptotic flatness condition with the help of the following recurrence relations (we use the notation of the Toy model in the Introduction):
	
	\begin{prop} \label{prop:recursion Klein-Gordon}
		Let $(M,g, \phi)$ be an asymptotically flat solution of the Einstein-massless-Klein-Gordon equations. Then the asymptotic quantities satisfy the following recurrence relations:
			\begin{subequations} 
	\begin{align}
	 \overset{(n+1)}{\alpha_{ij}} &= (n-1) \overset{(n)}{\chi_{ij}} - 2 \overset{(n)}{x} \overset{(1)}{x}  \eta_{ij} + \lfloor n-1 \rfloor , 	\\
	 \overset{(n)}{{h_i}^j} &= \overset{(n)}{{\chi_i}^k} \overset{(1)}{{h_k}^j} + \lfloor n-1 \rfloor , 
	\end{align}
	\end{subequations}

	\begin{subequations}
	\begin{align}
	 \overset{(n+1)}{\beta_i}, \overset{(n)}{\cancel{X}}, \overset{(n)}{\omega_{jji}}, \overset{(n)}{\zeta_i}, \overset{(n)}{f^i} &= \{ \overset{(n)}{\chi_{ij}}, \overset{(n)}{x}, \lfloor n-1 \rfloor \} ,	\quad  n> 3	
	\end{align}
	\end{subequations}

	\begin{subequations}	
	\begin{align}
	 \overset{(n+1)}{\rho}, \overset{(n+1)}{\si}, \overset{(n)}{\underline{x}}, \overset{(n)}{\underline{\omega}}, \overset{(n)}{\omega_{123}}, \overset{(n)}{\underline{\chi}_{ij}}, \overset{(n)}{\la_i},  \overset{(n-1)}{f^0} &= \{ \overset{(n)}{\chi_{ij}}, \overset{(n)}{x}, \lfloor n-1 \rfloor \} , \quad n> 2 .	
	 \end{align}
	\end{subequations}
Moreover, 
	\begin{subequations}	
	\begin{align}
	2\partial_u \overset{(n+1)}{\chi_{ij}} &= -n \overset{(n)}{\underline{\chi}_{ij}} + \{ \overset{(n)}{\chi_{ij}}, \overset{(n)}{x}, \lfloor n-1 \rfloor \} , 		\\
	2\partial_u \overset{(n+1)}{x} &= \{ \overset{(n)}{x}, \lfloor n \rfloor \}, 
	\end{align}
	\end{subequations}
	
	\begin{subequations} 
	\begin{align}
	\overset{(n+1)}{\underline{\be}_i} &= \partial_u \overset{(n+1)}{\zeta_i} + \lfloor n \rfloor , 			\\
	\overset{(n+1)}{\underline{\al}_{ij}} &= \partial_u \overset{(n+1)}{\underline{\chi}_{ij}} + \lfloor n \rfloor .
	\end{align}
	\end{subequations}
	\end{prop}
	
	In particular, if $\Xi = \overset{(1)}{\underline{\hat{\chi}}}$ and $\overset{(2)}{x} = -\overset{(1)}{\phi}$ vanish, then all the asymptotic quantities are $u$-independent.	\\
	
	Next, the proof Proposition \ref{prop:unique continuation2} works in exactly the same way as before to obtain:
	\begin{prop}	\label{prop:Klein-Gordon}
	Let $(M,g,\phi)$ be an asymptotically flat solution of the Einstein-massless-Klein-Gordon equations with rapidly decaying curvature in the sense that uniformly with respect to an orthonormal frame
	\begin{align} 
	\begin{aligned}
		|C_{\al \be \ga \de}| & = \order(r^{-3}),  \\	|\nabla_\ep C_{\al \be \ga \de}| &=\order(r^{-4}),	
	\end{aligned}
	&&
	\begin{aligned}
		|\nabla_{\al} \phi| &=\order(r^{-2}),  \\ 	| \nabla_{\al} \nabla_{\be} \phi| &=\order(r^{-3}).
	\end{aligned}
	\end{align}
Let $(u,s, \theta^2,\theta^3)$ be coordinates on $D_{\omega}$, $\omega>0$, constructed as in Section \ref{section:tetrad and coordinates} with $L:=\partial_s$ and $T:=\partial_u$ satisfying 
	\[	\nabla_L L=0 , \qquad	[L,T]=0	. \]
Assume that $T$ is a symmetry to all orders at infinity, i.e., for all $N\in \N$ there is an exhaustion $(\mathcal{U}_k)$ such that
	\begin{align}	
		\lim_{k\rightarrow \infty} \int_{\partial \mathcal{U}_k} r^N \Lie_T g = 0, 	\qquad
		\lim_{k\rightarrow \infty} \int_{\partial \mathcal{U}_k} r^N \Lie_T C = 0, 	\qquad
		\lim_{k\rightarrow \infty} \int_{\partial \mathcal{U}_k} r^N \Lie_T \phi = 0.
	\end{align}
Then $T$ is in fact a genuine symmetry for $(M,g,F)$, namely
	\[	\Lie_Tg \equiv 0, \quad \Lie_T C \equiv 0, \quad \Lie_T \phi \equiv 0, 	\quad \textrm{ on }	 \mathcal{D}_{\omega'}\]
	for some $0<\omega'<\omega$.
\end{prop}

\textbf{Sketh of proof.} Recall that $X_a:= \nabla_a \phi$ and let $Y:= \Lie_T X - B \odot X$ be the deformation tensor associated to $\phi$. \\

The wave equations obeyed by the deformation tensors in this case are simpler. Indeed we have that
		\begin{align*}
		\square W &= (R+C) \cdot W + (C^2 + X^2\cdot C + R\cdot C)\cdot B + (X \cdot C) \cdot Y + \nabla C \cdot \nabla B + C\cdot \nabla P, \\
		\square Y &=  \nabla X \cdot \nabla B + X \cdot \nabla P ,
		\end{align*}		
where we have used the fact that Klein-Gordon equation for a massless field, $\square \phi = 0$, implies $\square X = 0$ and Lemma \ref{lem:commutator} about commuting Lie and covariant derivatives. $_{\blacksquare}$ \\
\\
\textbf{Remark.} Note that the above system lacks the troublesome slow decaying terms. So in fact the proof of Theorem \ref{thm:Klein-Gordon} is completely analogous to the Alexakis-Schlue argument. Therefore the techniques employed in this paper allow to generalise Theorem \ref{thm:Klein-Gordon} to wave maps coupled directly to the Riemann/Weyl tensor. 

\section{Tetrad formalism} 

We give here a brief review of tetrad methods. We compare and contrast between the two main notations appearing in the literature, that is, a dictionary is presented between the Newman-Penrose spin coefficients and the Christodoulou-Klainerman null components. In the following Greek indices are used to enumerate the elements of a basis as well as the components of a tensor with respect to that basis. Einstein summation convention is used throughout: repeated indices are to be understood as summed over the range $0,1, \ldots, n$. \\

Let $(\mathcal{M}, \langle \cdot, \cdot \rangle )$ be an $(n+1)$-Lorentzian manifold, the signature convention will be $(-,+,\ldots,+)$. Let $\{e_{\mu}\} = \{e_0, e_1,\ldots,e_n\}$ be a basis of smooth vector fields on an open subset of $\mathcal{M}$. We will always assume that they form a null-orthonormal frame, that is, their dot product is given by
	 \[ \langle	e_{\mu}, e_{\nu} \rangle = \eta_{\mu \nu}	, \]
where 
			\begin{eqnarray} \label{metric-matrix}
	 \eta_{\mu\nu} = \left( \begin{array}{ccc}
	0 & -2 &  0 \\
	-2 & 0 &  0	\\
	 0 & 0 & \textrm{Id}_{n-1}\\
 \end{array}\right) .
\end{eqnarray}
It will be convenient to denote by $\{e^0, e^1, \ldots, e^n \}$ the basis of 1-forms dual to $\{e_0, e_1, \ldots, e_n\}$, that is, $e^{\mu}$ is the 1-form defined by $e^{\mu}(e_{\nu}) = {\delta^{\mu}}_{\nu}$. Using this notation the components of a tensor $T$ of type $(r,s)$ with respect to this basis are
	\[	 {T_{\mu...\nu}}^{\rho...\sigma} = T(e_{\mu},..., e_{\nu}, e^{\rho}, ..., e^{\sigma}) . \]
Then the tensor can be recovered from its components as follows,
		$$
	T  = {T_{\mu \ldots \nu}}^{\rho \ldots \sigma}  e^{\mu} \otimes \ldots \otimes e^{\nu} \otimes e_{\rho}\otimes  \ldots  \otimes e_{\si} .
		$$
Indices will be lowered and raised using the matrix $\eta_{\mu\nu}$ and its inverse, $\eta^{\mu \nu}$. Compatibility with the abstract-index lowering and raising operation is ensured by the relation
	$$ e^{\mu} = \langle \eta^{\mu \nu} e_{\nu}, \cdot	\rangle $$
which can be checked by direct evaluation. 	\\

The connection coefficients\footnote{Also known as Ricci or spin coefficients. Here we do not use that name to avoid confusion with the components of the Ricci tensor.} are the components of the derivative operator, 
	\begin{eqnarray*}
		\omega_{\lambda \mu \nu} = \langle \nabla_{e_{\lambda}} e_{\nu}, e_{\mu}	\rangle .
	\end{eqnarray*}
They satisfy $\omega_{\lambda \mu \nu} = -\omega_{\lambda \nu \mu}$, this is a consequence of $\eta_{\mu \nu}$ being a constant matrix.

\subsection{Frame equations and gauge conditions} 

Given a choice of coordinates $(x^0, x^1, \ldots, x^n)$ let $\{ \partial_{x^0}, \partial_{x^1}, \ldots, \partial_{x^n} \} $ be the associated basis. Recall that we have denoted by ${h_{\mu}}^{a}$ the \emph{orthonormalisation matrix}, that is, it is the change of basis defined by
	\[ e_{\mu} = {h_{\mu}}^{a} \partial_{x^{a}}	. \]
We have made emphasis on the fact that $a$ refers to an enumeration of the coordinate basis as opposed to the frame basis. In particular we have to be careful when contracting ${h_{\mu}}^{a}$ with the components of a tensor; the upstairs index only eats components with respect to the coordinate basis whereas the downstairs index only eats components with respect to the null-orthonormal frame. \\

The \textit{frame equations} are PDEs relating the orthonormalisation matrix components with the connection coefficients: 

\begin{lem}\emph{The frame equations.} The following equations hold
	\begin{eqnarray} \label{frame}
		e_{\mu} ({h_{\nu}}^{a}) - e_{\nu} ({h_{\mu}}^{a}) =  ({\omega_{\mu}}^{\rho} {}_{\nu} - {\omega_{\nu}}^{\rho} {}_{\mu}) {h_{\rho}}^{a} .
	\end{eqnarray}
\end{lem} 
\textbf{Proof.} This is precisely the torsion-free property of the connection, 
	\[ [e_{\mu}, e_{\nu}] = \nabla_{e_{\mu}} e_{\nu} - \nabla_{e_{\nu}}e_{\mu}.	\]
	Applying this to the coordinate function $x^{a}$ we get, 
	\begin{eqnarray*}
		[e_{\mu}, e_{\nu}] (x^{a}) &=& 2 e_{[\mu} (e_{\nu]}(x^{a} ) ), \\
										 &=& 2 e_{[\mu} ({h_{\nu]}}^{\rho} \partial_{x^{\rho}}(x^{a} ) )	,\\
										 &=& 2 e_{[\mu} ({h_{\nu]}}^{a} ) .
	\end{eqnarray*}
	On the other hand,
	\begin{eqnarray*}
		[e_{\mu}, e_{\nu}] (x^{a}) &=& ( \nabla_{e_{\mu}} e_{\nu} - \nabla_{e_{\nu}}e_{\mu} ) (x^{a} ) , \\
										&=& ({\omega_{\mu}}^{\rho} {}_{\nu} - {\omega_{\nu}}^{\rho} {}_{\mu}) e_{\rho} (x^{a}), 	\\
										&=& ({\omega_{\mu}}^{\rho} {}_{\nu} - {\omega_{\nu}}^{\rho} {}_{\mu}) {h_{\rho}}^{b}	 \partial_{x^{b}} (x^{a}), 	\\
										&=& ({\omega_{\mu}}^{\rho} {}_{\nu} - {\omega_{\nu}}^{\rho} {}_{\mu}) {h_{\rho}}^{b} {\delta_{b}}^{a} , 	\\
										&=& ({\omega_{\mu}}^{\rho} {}_{\nu} - {\omega_{\nu}}^{\rho} {}_{\mu}) {h_{\rho}}^{a} . \hspace{1cm}		 _{\blacksquare}
	\end{eqnarray*}

Now, recall the coordinates and tetrad constructed in Section \ref{section:tetrad and coordinates}. We will assume for the time being that $e_0=L=\partial_s$ is only parallel to the degenerate direction of the null hypersurfaces $C_u^+$, that is, we do not require it to be geodesic. 

\begin{lem} \label{lem:null gauge}
Suppose we have coordinates $(s,u, x^2, x^3)$ such that the level sets of $u$ are null hypersurfaces whose degenerate direction is parallel to $\partial_s$ and define the tetrad $\{e_0, e_1, e_2,e_3\}$ as in (\ref{tetrad}). Then we have that $\chi_{ij}$ and $\underline{\chi}_{ij}$ are symmetric and $\omega_{00i}=0$. Moreover, the $s$-curves are pre-geodesic and they can be reparametrised to be geodesics; this latter condition is equivalent to $\omega_{001}=0$.
\end{lem}
\textbf{Proof.} The symmetry of $\chi_{ij}$ follows from the ${_{ij}}^{u}$-component of the frame equations and the fact that ${h_i}^{u}=0$, $i=2,3$. Similarly for $\underline{\chi}_{ij}$. \\

The vanishing of $\omega_{00i}$ follows from the ${_{0i}}^{u}$-component of the frame equations and the fact that the first column of ${h_{\mu}}^{a}$ is constant. Here we include a less obscure computation which in addition clarifies the relation to the null foliation. Indeed, note that 
	\begin{align*}
		\omega_{00i} & = \langle \nabla_{e_0} e_i , e_0	\rangle = \langle [e_0,e_i] + \nabla_{e_i} e_0 , e_0	\rangle	.
	\end{align*}
Now, given the choice of tetrad, the vector fields $e_0$ and $e_i$ lie in the tangent space of $C_u^+$, then so does its commutator $[e_0,e_i]$. Moreover, $e_0$ is the degenerate direction in that tangent space, hence $ \langle [e_0,e_i] , e_0	\rangle =0$. On the other hand $\langle  \nabla_{e_i} e_0 , e_0	\rangle	= \frac12 e_i \langle e_0 , e_0	\rangle = 0	$. Therefore  $\omega_{00i}=0$.\\

The previous computation implies that the vector field $\nabla_{e_0}  e_0$ is orthogonal to $C_u^+$, thus it must parallel to the degenerate direction, that is $e_0$. Hence $e_0$ is pre-geodesic. Finally we note that $\omega_{010}=\langle \nabla_{e_0} e_0 , e_1	\rangle	$ is precisely the obstruction to $e_0$ being exactly geodesic. $_{\blacksquare}$	\\

Now we will see that with a judicious choice of ${h_{i}}^j$ we can achieve more cancellations. 

\begin{lem} \label{lem:comoving gauge}
	The orthonormal basis $\{ e_i : i=2,  \ldots, n \}$ on the surfaces $S_{s,u}$ can be chosen such that $\omega_{0ij} = 0$.
\end{lem}
\textbf{Proof.} The idea is that $\omega_{0ij}$ consists of $(n-1)(n-2)/2$ algebraically independent components. This coincides with the number of degrees of freedom of orthogonal symmetries on a $(n-1)$-plane, that is, dim $SO(n-1) = (n-1)(n-2)/2$. So we can solve a system of ODEs to achieve $\hat{\omega}_{0ij} = 0$. \\

Explicitly, under a rotation ${\Theta_i}^j\in SO(n-1)$,
	\[ e_i \mapsto	\hat{e}_i = {\Theta_i}^j e_j ,	\]
the quantities $\omega_{0ij}$ transform to $\hat{\omega}_{0ij}$ where
	\begin{align*}
		\hat{\omega}_{0ij} &= \langle \nabla_{e_0} \hat{e}_j, \hat{e}_i	\rangle , 	\\
							&= \langle \nabla_{e_0} {\Theta_j}^l e_l, {\Theta_i}^k e_k	\rangle , 	\\
							&= {\Theta_i}^k  ( {\Theta_j}^l \omega_{0kl} + e_0({\Theta_j}^l ) \delta_{kl} ).
	\end{align*}
Thus, by solving $ {\Theta_j}^l \omega_{0il} + e_0({\Theta_j}^l ) \delta_{il} = 0$ (recall that we have the correct number of equations and variables) we can set $\hat{\omega}_{0ij} = 0$ and omit the hat hereafter. ${}_{\blacksquare}$	\\

The previous argument is invariant under rescaling of $e_0=\partial_s$. We know from Lemma \ref{lem:null gauge} that any null vector field tangent to a null foliation is pre-geodesic, that is, its flow-lines can be reparametrised to be geodesics. Therefore we can achieve the null-geodesic and co-moving gauges simultaneously. 

\subsection{Christodoulou-Klainerman and Newman-Penrose notations} \label{NP-CK notation}

Here we introduce the Christodoulou-Klainerman null components of the connection and the Weyl and Faraday tensors. Also a table is presented comparing the slight variations adopted in this paper and the Newman-Penrose spin coefficients.	\\

We will work with the convention that $e_0$ is to be thought as the out-going null direction while $e_1$ corresponds to the in-going null direction. Hereafter, Latin indices $i$, $j$, $\ldots$, will run from 2 to 3. With this in mind we define the Christodoulou-Klainerman null components of the connection\footnote{They use the last indices to refer to the null part of the frame, that is, their null pair $(e_3,e_4)$ corresponds to our $(e_1,e_0)$.} by, 
	\begin{align*}
		\chi_{ij} &:= \langle \nabla_{e_i} e_0, e_j \rangle = \omega_{ij0} ,   \qquad	
					&&	\underline{\chi}_{ij} := \langle \nabla_{e_i} e_1, e_j \rangle = \omega_{ij1} ,  \\
		2 \xi_i &:= \langle \nabla_{e_0} e_0, e_i \rangle =  \omega_{0i0} ,   \qquad	
					&&	2 \la_i := \langle \nabla_{e_1} e_1, e_i \rangle = \omega_{1i1} ,  \\
		2 \zeta_i &:= \langle \nabla_{e_1} e_0, e_i \rangle = \omega_{1i0} ,   \qquad	
					&&	2 \underline{\zeta}_i := \langle \nabla_{e_0} e_1, e_i \rangle = \omega_{0i1} ,  \\
		4\omega &:= \langle \nabla_{e_0} e_0, e_1 \rangle = \omega_{010} , 	\qquad	
					&&	4 \underline{\omega} := \langle \nabla_{e_1} e_1, e_0 \rangle = \omega_{101} , 	\\
		V_i & := \langle \nabla_{e_i} e_0, e_1 \rangle = \omega_{i10}.
	\end{align*}
The Weyl curvature null components are given by,
	\begin{align*}
		\alpha_{ij} &:= C_{i0j0} ,   \qquad	&&	\underline{\alpha}_{ij} := C_{i1j1} ,  \\
		2 \beta_i &:= C_{i010} ,   \qquad	&&	2 \underline{\beta}_i := C_{i110} ,  \\
		4 \rho &:= C_{1010} , 	\qquad	&&	2 \sigma := C_{1023} .	
	\end{align*}
Due to the symmetries of the Weyl tensor we have that $\al_{ij}$ and $\underline{\alpha}_{ij}$ are trace-less and symmetric. Moreover, in $(3+1)$-dimensions the above components determine completely the Weyl tensor, in particular we have
	\begin{align*}
		2 \beta_i &= C_{i010} = 2C_{0jji} , \quad i\neq j ,  \quad	&	2 \underline{\beta}_i &= C_{i110} = -2C_{1jji},  \quad i\neq j , \\
		4 \rho &= C_{1010} = -C_{0212}  	\quad	&	2 \sigma &= C_{1023} = 2C_{1203} 		\\
				& = -C_{0313}=-C_{2323} , 	\quad	&            &= 2C_{0312} = -2C_{0213} .
	\end{align*}
Finally, given a Faraday tensor, its null components are defined as:
\begin{align*}
	\alpha(F)_i &= F_{i0} ,			\quad& 	\underline{\alpha}(F)_i &= F_{i1} ,	\\
	\rho (F)	&= \frac12F_{10},	\quad&		\sigma(F) &= F_{23} .
\end{align*}
These components determine completely the Faraday tensor. \\ 

In order to compare with the Newman-Penrose notation, \cite{NP}, we set, 
	\[ l = \frac{1}{\sqrt{2}}e_0, \quad n=\frac{1}{\sqrt{2}} e_1 , \quad m = \frac{1}{\sqrt{2}} (e_2 + i e_3). \]
Table 1 summarises the correspondence between the different notations. Also the concept of signature explained in Section \ref{section:tetrad and coordinates} is included. 

\begin{table}  \label{table:CK-NP}  
\caption{Comparison of different notations.}	
\centering 

\begin{tabular}{ |c| c| c| c|}		
 \hline                     
CK	&	NP		& This paper	&  Signature		\\
\hline	\hline
 $L $	&	$l$	&	$e_0$	& 1 \\
 $\underline{L} $	&	$n$	&	$e_1$	&	-1 \\
 $e_i $	&	$m$	&	$e_i$	&  0 \\
\hline
$\xi_i$ &	$\kappa$	&	$\omega_{0i0}$		&	2		\\
$\hat{\chi}_{ij}$, $\tr \chi$ &	$\si$, $\rho$	&	$\chi_{ij}=\omega_{ij0}$		&	1	\\
$\omega, \cancel{\nabla}_{L}e_i$ &	Re $\ep$ , Im $\ep$	&	$\omega_{010}$, $\omega_{023}$		&	1	\\
$\zeta_i$ &	$\tau$	&	$\omega_{1i0}$		&	0					\\
$\underline{\zeta}_i$ &	$\pi$	&	$\omega_{01i}$		&	0			\\
$V_i$ &	$\bar{\al}+\beta$	&	$\omega_{i10}$		&	0				\\
$\cancel{\nabla}_{e_i}e_j$ &	$\bar{\al}-\beta$	&	$\omega_{223}$, $\omega_{332}$ & 0	\\
$\hat{\underline{\chi}}_{ij}$, $\tr \underline{\chi}$ &	$\la$, $\mu$	&	$\underline{\chi}_{ij}=\omega_{ij1}$		&	-1	\\
$\underline{\omega}$, $\cancel{\nabla}_{\underline{L}}e_i$ &	Re $\ga$ , Im $\ga$	&	$\omega_{101}$, $\omega_{123}$ & -1	\\
$\underline{\xi}_i$ &	$\nu$	&	$\la_i=\omega_{1i1}	$	&	-2		\\
\hline 
$\alpha_{ij}$	&	$\Psi_0$	&	$C_{i0j0}$	&	2	\\
$\beta_i$	&	$\Psi_1$		&	$C_{i010}$	&	1	\\
$\rho$, $\si$	&	$\Psi_2$	&	$C_{1010}$, $C_{1023}$	&	0	\\
$\underline{\beta}_i$	&	$\Psi_3$		&	$C_{i110}$	&	-1	\\
$\underline{\alpha}_{ij}$	&	$\Psi_4$	&	$C_{i1j1}$	&	-2	\\
\hline
-	&	$\Phi_{00}$	&	$S_{00}$	&	2	\\
-	&	$\Phi_{01}$		&	$S_{0i}$	&	1	\\
-	&	$\Phi_{11}$, $\Lambda$, $\Phi_{02}$	&	$S_{01}$, $S_{22}$, $S_{23}$, $S_{33}$	&	0	\\
-	&	$\Phi_{12}$		&	$S_{1i}$	&	-1	\\
-	&	$\Phi_{22}$	&	$S_{11}$	&	-2	\\
\hline
$\alpha_i(F)$	&	$\phi_0$	&	$F_{0i}$	&	1	\\
$\rho(F)$, $\si(F)$	&	$\phi_1$	&	$F_{01}$, $F_{23}$	&	0	\\
$\underline{\alpha}_{i}(F)$	&	$\phi_2$	&	$F_{1i}$	&	-1	\\
\hline
\end{tabular}
\end{table}

\newpage

\bibliography{mybib}
\bibliographystyle{plain}

\end{document}